\documentclass[aps,prd,showpacs,twocolumn,amssymb,floatfix]{revtex4}
\usepackage{graphicx}
\usepackage{longtable}

\usepackage{amsmath, amsthm, amssymb}
\usepackage{color}
\def\lsim{\mathrel{\rlap{
\lower4pt\hbox{\hskip-3pt$\sim$}}
\raise1pt\hbox{$<$}}}     
\def\gsim{\mathrel{\rlap{
\lower4pt\hbox{\hskip-3pt$\sim$}}
\raise1pt\hbox{$>$}}}     

\def\be{\begin{eqnarray}}
\def\ee{\end{eqnarray}}
\def\prt{\partial}

\begin{document}
\title{Electromagnetic field evolution in relativistic heavy-ion collisions}

\author{V. Voronyuk}
\affiliation{ Joint Institute for Nuclear Research,  Dubna,
Russia}
\affiliation{Frankfurt Institute for Advanced Study, Frankfurt,
Germany}
\affiliation{Bogolyubov Institute for Theoretical Physics, Kiev,
Ukraine}

\author{V. D. Toneev}
\affiliation{ Joint Institute for Nuclear Research,  Dubna, Russia}
\affiliation{Frankfurt Institute for Advanced Study, Frankfurt,
Germany}

\author{W. Cassing}
\affiliation{Institute for Theoretical Physics, University of
Giessen, Giessen, Germany}

\author{E. L. Bratkovskaya}
\affiliation{Institute for Theoretical Physics, University of
Frankfurt, Frankfurt, Germany }
\affiliation{Frankfurt Institute for Advanced Study, Frankfurt,
Germany}

\author{V. P. Konchakovski}
\affiliation{Institute for Theoretical Physics, University of
Giessen, Giessen, Germany}
\affiliation{Bogolyubov Institute for Theoretical Physics, Kiev,
Ukraine}

\author{S. A. Voloshin}
\affiliation{Wayne State University, Detroit, Michigan, USA  }

\begin{abstract}
The hadron string dynamics (HSD) model is generalized to
include the creation and evolution of retarded electromagnetic fields
as well as the influence of the magnetic and electric fields
on the quasiparticle propagation. The time-space structure of the
fields is analyzed in detail for non-central Au+Au collisions at
$\sqrt{s_{NN}}=$200 GeV. It is shown that the created magnetic field
is highly inhomogeneous but in the central region of the overlapping
nuclei it changes relatively weakly in the transverse direction.
For the impact parameter $b=$10 fm the maximal magnetic field -
perpendicularly to the reaction plane - is obtained  of order
$eB_y/m_\pi^2\sim$5 for a very short time $\sim$ 0.2 fm/c, which
roughly corresponds to the time of a maximal overlap of the
colliding nuclei.  We find that at any time the location of the
maximum in the $eB_y$ distribution correlates with that of the energy
density of the created particles. In contrast, the electric field
distribution, being also highly inhomogeneous, has a minimum in the
center of the overlap region. Furthermore, the field characteristics
are presented as a function of the collision energy and the
centrality of the collisions.  To explore the effect of
the back-reaction of the fields on hadronic observables a
comparison of HSD results with and without fields is exemplified.
 Our actual calculations show  no noticeable influence of the
electromagnetic fields - created in heavy-ion collisions - on the
effect of the electric charge separation with respect to the 
reaction plane.
\end{abstract}

\maketitle

\section{Introduction}

The study of the properties of QCD matter in the presence of strong
uniform magnetic fields has attracted much attention during 
recent years due to several remarkable observations. 
They include such a universal phenomenon as the magnetic
catalysis~\cite{KGMS92,KZ08,PRA10} in which the magnetic field acts
as a strong catalyst of dynamical flavor symmetry breaking which
might lead to the generation of a dynamical fermion mass.
Furthermore, in dense QCD matter  in the presence of an external
magnetic field and/or topological defects, a  spontaneous creation of
axial currents~\cite{SZ04} may happen. Moreover, at finite baryon
density, due to a response of the QCD ground state to a strong
magnetic field, a metastable object, the $\pi^0$ domain wall (or
``Goldstone current state'' in quark matter), could appear which
energetically may be more favorable than nuclear matter at the same
density~\cite{SS07}. The presence of a magnetic field also favors
the formation of spatially inhomogeneous spiral-like quark condensate
configurations at low temperatures and  non-zero  chemical
potentials~\cite{FZK07}. The influence of a constant magnetic field
on possible color-superconducting phases (the color Meissner effect) has
also actively been discussed~\cite{CsupC}.  A clarification of
such phenomena by experimental observations requires e.g. the
production of QCD matter in relativistic heavy-ion collisions where
in non-central reactions strong electromagnetic fields are created
by the charged four-current of the spectators.

In this general respect exact solutions of the quantum field
equations of motion are of special interest. The latter supply us
with microscopic insight for problems of relativistic charged
particle motion in electromagnetic fields of terrestrial
experimental devices as well as in astrophysics and cosmology. In
particular, they apply to the  development of the quantum theory of
synchrotron radiation~\cite{ST68} and also for the description of
interacting particles, including electrons and neutrinos, especially
in matter in the presence of external electromagnetic
fields~\cite{TKTH10}.

 More specifically, the  quark-hadron and chiral
symmetry restoration transitions might be dramatically modified in
the presence of a strong magnetic field.  We recall that
without a magnetic field a crossover at $T = T_c \approx$160 MeV is
realized at vanishing baryon chemical potential $\mu_q$. The
presence of a strong magnetic background turns this picture within a
linear sigma model into a weak first order phase
transition~\cite{AF08,FM08} whereas in the Nasmbu-Jona-Lasinio 
NJL model the crossover
remains ~\cite{BB10,CMM11}. This contradiction is reconciled by
noting a crucial role of of the vacuum contribution from quarks.
The vacuum contribution seems to soften the order of phase transition:
the first order phase transition - which would be realized in the absence 
of the vacuum contribution  - becomes a smooth crossover if the system with
vacuum quark loops included~\cite{MCF10}. 

 Within the Polyakov-Nambu-Jona-Lasinio (PNJL model) it was shown that
the external field acts as a catalyzer for dynamical symmetry
breaking which increases the critical temperature with increasing
strength of the magnetic field  and sharpens the transition
\cite{FRG10} in agreement with lattice QCD calculations~\cite{latB}. Very
recently, an astonishing feature has been demonstrated in effective
models in that lines of the finite temperature deconfinement and chiral
transitions can split in a strong magnetic background~\cite{MCF10,GR10}.
 We also mention  that the presence of a strong constant
magnetic field modified also the nature of the electroweak phase
transition in the evolution of the universe at its early
stages~\cite{EWtr}.

In the last few years,  particular attention has been paid to the
chiral magnetic effect (CME) closely related to a possible local
$\cal{P}$ and $\cal{CP}$ symmetry violation in strong interactions
as suggested in Ref~\cite{Kharzeev:2004ey} and widely
discussed in Refs.~\cite{KZ07,KMcLW07,FKW08,KW09,FRG10}.
 This effect originates from the existence of nontrivial topological
configurations of gauge fields and their interplay with the chiral
anomaly which results in an asymmetry between left- and right-handed
quarks. Such a chiral asymmetry, when coupled to a strong magnetic
field as created by colliding nuclei perpendicularly to the reaction
plane, induces an electric charge current  along the direction of a
magnetic field thereby separating particles of opposite charges with
respect to the reaction plane. Thus,  such topological effects in
QCD might be observed in heavy-ion collisions directly in the
presence of very intense external electromagnetic fields due to the
``Chiral Magnetic Effect''  as a manifestation of the spontaneous
violation of the $\cal {CP}$ symmetry.  Recently, topological charge
fluctuations and possible CME have been confirmed by QCD lattice
calculations  in quenched $SU(2)$ gauge theory~\cite{BCLP09} and in
QCD+QED with dynamical $2+1$ quark flavors~\cite{ABPZ09}.

One should note that in contrast to all the cases mentioned above,
the magnetic field in the CME is not constant and acts only during a
very short time.  The maximal strength of the electromagnetic field
$eB_y$ created in relativistic heavy-ion collisions is very high
$eB_y\sim$5 $m_\pi^2$ but its duration is $t\sim$0.2\ fm/c for Au+Au
at $\sqrt{s_{NN}}=$200\ GeV collisions and impact parameter $b=$10\
fm,   as was  estimated dynamically in~\cite{KMcLW07,SIT09,TV10}.

It is remarkable that the proposed CME observable, i.e. the electric
charge asymmetry of produced particles with respect to the reaction
plane, has been recently measured by the STAR
Collaboration~\cite{Vol09,STAR-CME}. Qualitatively the experimental results
agree with the magnitude and gross features of the theoretical
predictions for local $\cal {P-}$odd violation in heavy-ion
collisions. The observed signal can not be described by the
background models used in~\cite{Vol09,STAR-CME}, however, alternative
mechanisms resulting in a similar charge separation effect are not
fully excluded (see e.g. Refs. \cite{Wa09,Pr10,BKL09,BKL10,AMM10,MS09}). 
This issue is under intensive debate now.

 The aim of this paper  is to study the space-time evolution of
(electro-)magnetic fields formed in relativistic heavy-ion
collisions. The Hadron String Dynamics (HSD) transport
code~\cite{HSD} is used as a basis of our considerations. In
contrast with the UrQMD model without including electromagnetic
fields (previously used for the analysis~\cite{Vol09,STAR-CME}), the HSD model
corresponds to Kadanoff-Baym rather than Boltzman kinetic equations
and treats the nuclear collisions in terms of quasiparticles with a
finite width. In our approach the dynamical formation of the
electromagnetic field, its evolution during a collision and
influence on the quasiparticle dynamics as well as the interplay of
the created magnetic and electric fields and back-reaction effects
are included simultaneously.

The article is organized as follows: In Sect. II the model is
presented and it is shown how the formation of the electromagnetic
field and particle propagation in this background can be implemented
in the HSD transport code. In Sect. III the space-time structure of
the formed electromagnetic  fields, their correlations with the
energy density of produced particles, the correlation between the
magnetic and electric fields and some estimates for the
dependence of these characteristics on collision energy and impact
parameter are investigated. A comparison of some observables, which
are calculated within the HSD model with and without the
electromagnetic fields, is presented in Sect. IV. The results are
summarized in Sect.~V.

\section{Electromagnetic field evolution within the HSD model}

To describe a collision of  heavy ions let us start from the
relativistic Boltzmann equation for an on-shell phase-space
distribution function $f\equiv f(x,p)$ \be p^\mu \
\frac{\partial}{\partial x^\mu}  f = C[f] \label{RBeq}\ee where
$C[f]$ is the collision integral and $x,p$ are the 4-coordinate and
4-momentum of a particle. A background electromagnetic field may be
taken into account by including an electromagnetic tensor
$F_{\mu\nu}$ into Eq.(\ref{RBeq})  as \be
p^\mu(\frac{\partial}{\partial x^\mu} -
F_{\mu\nu}\frac{\partial}{\partial p^\nu} ) f = C[f] \label{RBM},\ee
 where
 \be
eF_{\mu\nu} = \partial_\mu A_\nu - \partial_\nu A_\mu
 \label{EMten}\ee
 with the electromagnetic 4-vector potential $A_\mu=\{\Phi,{\bf A}\}$.
Note that the left-hand side of Eq.(\ref{RBM}) is gauge invariant.

Let us rewrite (\ref{RBM}) it terms of components of $A_\mu$
and generalize it to the case of a set of particles moving in a nuclear
potential $U$. Then the equation
for a test particle characterized by the distribution function $f\equiv
f({\bf r},{\bf p},t)$ may be presented as follows:
\be \label{EMf}
& &\left\{ \frac{\prt}{\prt t}+\left(\frac{\bf p}{p_0} +
 {\bf \nabla}_{\bf p} \ U \right) {\bf \nabla}_{\bf r} \right. \nonumber  \\
 &-&\left.\left( {\bf \nabla}_{\bf r} \ U  +e{\bf \nabla }\Phi
 +e\frac{\prt {\bf A}}{\prt t}-  e{\bf v}\times
 ({\bf \nabla} \times {\bf A})\right){\bf \nabla}_{\bf p} \
 \right\} \ f  \nonumber \\
&=&C_{coll}(f,f_1,...f_N)   \ee
The strength of the magnetic ${\bf B}$ and electric ${\bf E}$ fields is, 
respectively,
 \be \label{def}
 {\bf B}= {\bf \nabla} \times {\bf A}~,  \ \ \ \ \ {\bf E}=
-{\bf \nabla} \Phi -\frac{\prt {\bf A}}{\prt t}~.
 \ee
 One should note that the electromagnetic field generated by moving nuclei may be
considered as an external field: the value of the electromagnetic
field at a given point is determined by the global charge
distribution of colliding nuclei and thus, in good approximation,
independent of the local strong interaction dynamics. However, the
presence of the electromagnetic field can affect the interactions
among particles, which simultaneously carry electric and (possibly)
color charges.

Using relations (\ref{def}), the system (\ref{EMf}) is reduced to a
more familiar form:
 \be \label{kin1} &&\left\{ \frac{\prt}{\prt
t}+\left(\frac{\bf p}{p_0} +
 {\bf \nabla}_{\bf p} \ U \right) {\bf \nabla}_{\bf r} \ + \right.\\
 &&\left.\left( -{\bf \nabla}_{\bf r} \ U  +e{\bf E}+ e{\bf v}\times
 {\bf B} \right){\bf \nabla}_{\bf p} \
 \right\} f =C_{coll}(f,f_1,...f_N)  \nonumber
 \ee
for particles of a charge $e$.

The  HSD transport model is based on Kadanoff-Baym equations for 
Green's function accounting for the first order gradient expansion of
the Wigner transformed Kadanoff-Baym
equation~\cite{CJ99,CJ00_1,CJ00_2}.  The hadronic mean field in
Eq.(\ref{EMf}) is  $U\sim Re (\Sigma^{ret})/2p_0$ where $\Sigma^{ret}$ 
denotes the retarded selfenergy. The change of the
hadron mass in the HSD model results in an additional term ahead of
${\bf \nabla}_{\bf p}$ in (\ref{EMf})~\cite{Iv87} but it is ignored
in our consideration which focuses on electromagnetic effects.

Transport equations for a strongly interacting particle with
electric charge $e$ (Eq.\ref{EMf}) are supplemented by the
electromagnetic field equations \be \label{Max} {\bf \nabla} \times
{\bf E}= - \frac{1}{c}\frac{\prt {\bf B}}{\prt t}~, \ \ \ \ {\bf
\nabla \cdot} {\bf B}=0~. \ee

 The general solution of the wave equations (\ref{Max}) with the charge
distribution $\rho({\bf r},t)=en$ and electric current
 ${\bf j}({\bf r},t)=e{\bf v}$ are
 \be
\label{vecP}  \Phi({\bf r},t) = \frac{1}{4 \pi} \int
\frac{\rho({\bf r'},t')\  \delta(t-t'- |{\bf r} - {\bf r'}|/c)}{
|{\bf r} - {\bf r'}|}  \ d^3r ' dt'
 \ee
 for the electromagnetic scalar potential $\Phi({\bf r},t)$ and
 \be
\label{scP}  {\bf A}({\bf r},t) = \frac{1}{4 \pi} \int
\frac{{\bf j}({\bf r'},t') \ \delta(t-t'- |{\bf r} - {\bf r'}
|/c)}{ |{\bf r} - {\bf r'}|}  \ d^3r ' dt'
 \ee
\noindent
 for the vector potential.  For a moving point-like charge one gets
 \be
 \label{11.42}  \rho({\bf r},t) =
e \ \delta({\bf r} - {\bf r}(t)), \hspace{0.3cm} {\bf j}({\bf r},t)
= e \ {\bf v}(t)\ \delta({\bf r} - {\bf r}(t)) \ee and, after
integration of Eq.(\ref{scP})  using \be \label{11.45}
\int_{-\infty}^\infty  g(x) \ \delta(f(x))\ dx = \sum_i
\frac{g(x_i)}{|f'(x_i)|} , \ee
we obtain: \be \label{11.46}
\Phi({\bf r},t) = \frac{e}{4 \pi \epsilon_0} \sum_i \frac{1}{R(t_i')
\kappa(t_i')} \ee with the definitions
 \be \label{11.47}
  {\bf R}(t'_i) &=& {\bf r} - {\bf
r}(t'_i),  \\ \nonumber \kappa(t'_i) &=& 1 - \frac{{\bf R}(t'_i)  \cdot {\bf
v}(t'_i)}{c R(t'_i)} = \left|\left(\frac{d f}{dt'}\right)_{t'=t'_i}\right| \ .
\ee
In (\ref{11.47}) the times  $t_i'$ are solutions of the retardation equation
\be \label{tp}
f(t')= t' - t + R(t')/c = 0.
\ee
By analogy, for the vector potential we get:
\be \label{11.48} {\bf A}({\bf r},t) = \frac{e \mu_0}{4
\pi} \sum_i \frac{{\bf v}(t_i')}{R(t_i') \kappa(t_i')}
\ee
Thus, Eqs.(\ref{vecP}) and (\ref{scP}) lead to the retarded
Li\'enard-Wiechert  potentials (\ref{11.46}) and (\ref{11.48}) acting at 
the point ${\bf R}={\bf r}-{\bf r'}$ at the moment $t$. The electromagnetic
potentials $\Phi({\bf r}, t)$ and  ${\bf A}({\bf r},t)$ are
generated by every moving charged particle and describe generally
the elastic Coulomb scattering as well as inelastic bremsstrahlung
processes. Below we set $\varepsilon_0=\mu_0=$1 in Eqs.(\ref{11.46})
and (\ref{11.48}), respectively. The retarded electric and magnetic
fields can be derived now from Eqs.(\ref{11.46}) and (\ref{11.48})
using  (\ref{def}) as follows:

\begin{widetext}
\be \label{E1}
 {\bf E}({\bf r},t) &=&
\frac{e}{4 \pi } \left( \frac{{\bf n}}{\kappa
R^2} +  \frac{-{\bf
b}/c - [ ({\bf n} \cdot {\bf v}) {\bf n}
- {\bf v}]/R } { \kappa^2 c R}\right)_{ret}  \nonumber \\
 &-& \frac{e}{4 \pi } \left(\frac{(-{\bf
v}(t')/c + {\bf n}(t'))({v^2}/{c} - {\bf n} \cdot {\bf v} -{R}/{c} ({\bf n} \cdot
{\bf b}))}{\kappa^3 c R^2}   \right)_{ret}~,  \ee
\be \label{E2}
{\bf B}({\bf r},t) &=& \frac{ e}{4 \pi} \left( \frac{{\bf v}
\times {\bf n}}{\kappa R^2} +
 (\frac{{\bf b}(t') \times {\bf n}(t') + {\bf v}(t') \times
 [ ({\bf n} \cdot {\bf v}) {\bf n}
- {\bf v}]/R }{ \kappa^2 c  R}) \right)_{ret}  \nonumber \\
  &-& \frac{ e}{4 \pi} \left(
\frac{ ({\bf v}(t') \times {\bf n}(t'))({v^2}/{c} -
{\bf n} \cdot {\bf v} -{R}/{c} ({\bf n} \cdot
{\bf b}))  }{ \kappa^3 c R^2} \right)_{ret}
\ee
\end{widetext}
with the acceleration  ${\bf b} = d/dt' {\bf v}$ and the unit vector
${\bf n}={\bf R}/R$. After simplification and neglecting the
acceleration $\bf b$ we arrive at
 \be
 e\,\bf{E} &=&  \frac{{\it sign}(e) \: \alpha \
 {\bf R}(t)\,(1-v^2/c^2)}{\left[({\bf R}(t)\cdot{\bf v}/c)^2+R^2(t) (1-v^2/c^2)\right]^{3/2}}~,
 \\
 e\,\bf{B} &=& \frac{{\it sign} (e) \:  \alpha \
 [{\bf v}\times{\bf R}(t)]\,(1-v^2/c^2)}{c\left[({\bf R}(t)\cdot{\bf v}/c)^2+R^2(t)
 (1-v^2/c^2)\right]^{3/2}}
 \ee
where  in the left-hand side an additional charge $e$ is introduced to get the
electromagnetic fine-structure constant  $\alpha=e^2/4\pi=1/137$ in the 
right-hand site of these equations. The last expression is
reduced to the familiar form of the retarded Li\'enard-Wiechert equation 
for the magnetic field of a moving charge
\be \label{LWeq}
\mathbf{B}(\mathbf{r}, t) = \frac{ e}{4 \pi} \:
\frac{[\mathbf{v}\times \mathbf{R}]}{c R^3} \:
\frac{(1-v^2/c^2)}{[1-(v/c)^2 \sin^2 \phi_{R v}]^{3/2}},
\ee
with $\mathbf{R}=\mathbf{R}(t)$, $\phi_{R v}$ is the angle between
$\mathbf{R}(t)$ and $\mathbf{v}$.

The set of transport equations (\ref{EMf}) in the following is
solved in a quasiparticle approximation by using the Monte-Carlo
parallel ensemble  method. To find the electromagnetic field, a
space grid is used. The quasiparticle propagation in the electromagnetic field
is calculated according to Eq.(\ref{kin1}) as 
\be \label{mom}
\frac{d{\bf p}}{dt}=
e{\bf E}+ \frac{e}{c}{\bf v}\times {\bf B}~. 
\label{ener}
 \ee
The change of the electromagnetic energy  is $(e/c)({\bf v\cdot E})$ {\it i.e.}
the magnetic field does not change the quasiparticle energy. To avoid singularities 
and self-interaction effects, particles within a given Lorentz-contracted cell 
are excluded from   the field calculation.

\section{Results}

\begin{figure}[thb]
\centering{\includegraphics[width=7.0truecm,clip=] {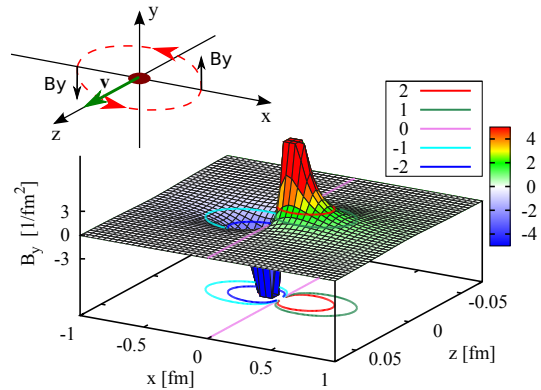} }
\caption{(Color online) Snapshot of the $B_y$ distribution for the magnetic field and
its projection on the $(z-x)$
plane for a single charge moving along the $z$ axis.
 }
\label{oneCh}
\end{figure}

The  scheme described above for the computation of the
electromagnetic field is applied here to ultra-relativistic
heavy-ion collisions. However,  for transparency, we shall start
with the magnetic field created by a single freely moving charge. As
seen in Fig.~\ref{oneCh},  the charge creates a cylindrically
symmetric field with the symmetry axis along the direction of
motion. If one follows the magnetic field direction, it appears to be
 torqued around the direction of motion. Therefore, the
magnetic field on the left and right sides with respect to the
moving charge  has  opposite signs, resulting in some maximum and
minimum of the magnetic field at a given instant of time. The
opposite field signs directly follow from Eq.(\ref{LWeq}) if one
takes into account that the vector ${\bf R}$ (Eq.(\ref{11.47})) in
this situation has opposite signs.

\begin{figure}[thb]
\includegraphics[width=5.0truecm,clip=] {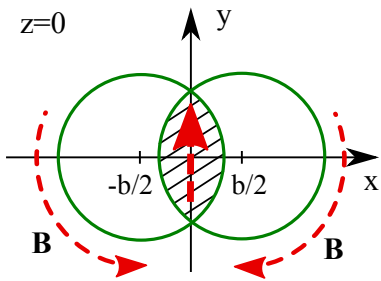}
\caption{(Color online) The transverse plane of a noncentral heavy-ion collision.
The impact parameter of the collision is denoted by $b$. The magnetic
field is plotted by the dashed lines. } \label{tr-pl}
\end{figure}

In  a nuclear collision, the magnetic field will be a
superposition of solenoidal fields from different moving charges. The collision
geometry for a peripheral collision  is shown in Fig.~\ref{tr-pl} in
the transverse plane. The overlapping strongly interacting region
(participants) has an ``almond''-like
shape. The nuclear region outside this almond (shaded 
in Fig.~\ref{tr-pl}) corresponds to
spectator matter which is the dominant source of the electromagnetic
field at the very beginning of the nuclear collision. Note that in
the HSD code the particles are subdivided into  target and projectile 
spectators and participants not geometrically but dynamically: 
spectators are nucleons which suffered yet no collision.

\subsection{Space-time evolution of the magnetic field}
\begin{widetext}

\begin{figure*}[thb]
\includegraphics[height=6.0truecm] {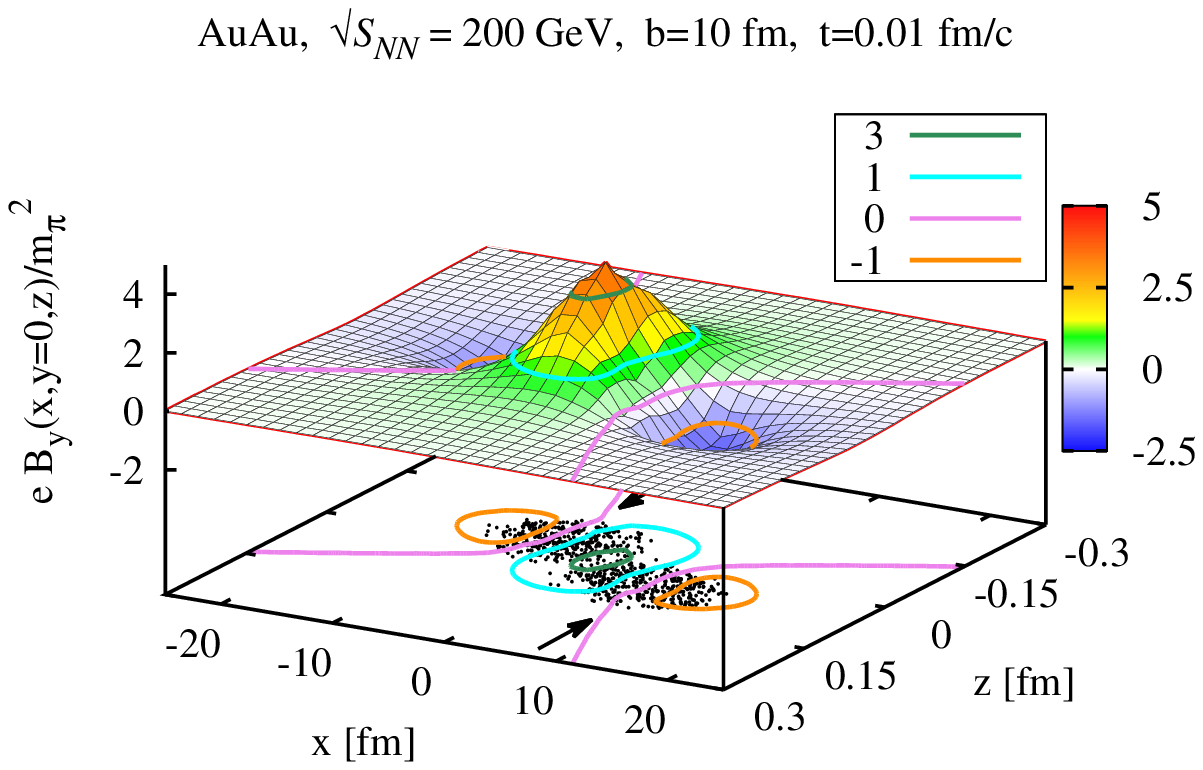}
\hspace{0.1cm}
\includegraphics[height=6.0truecm] {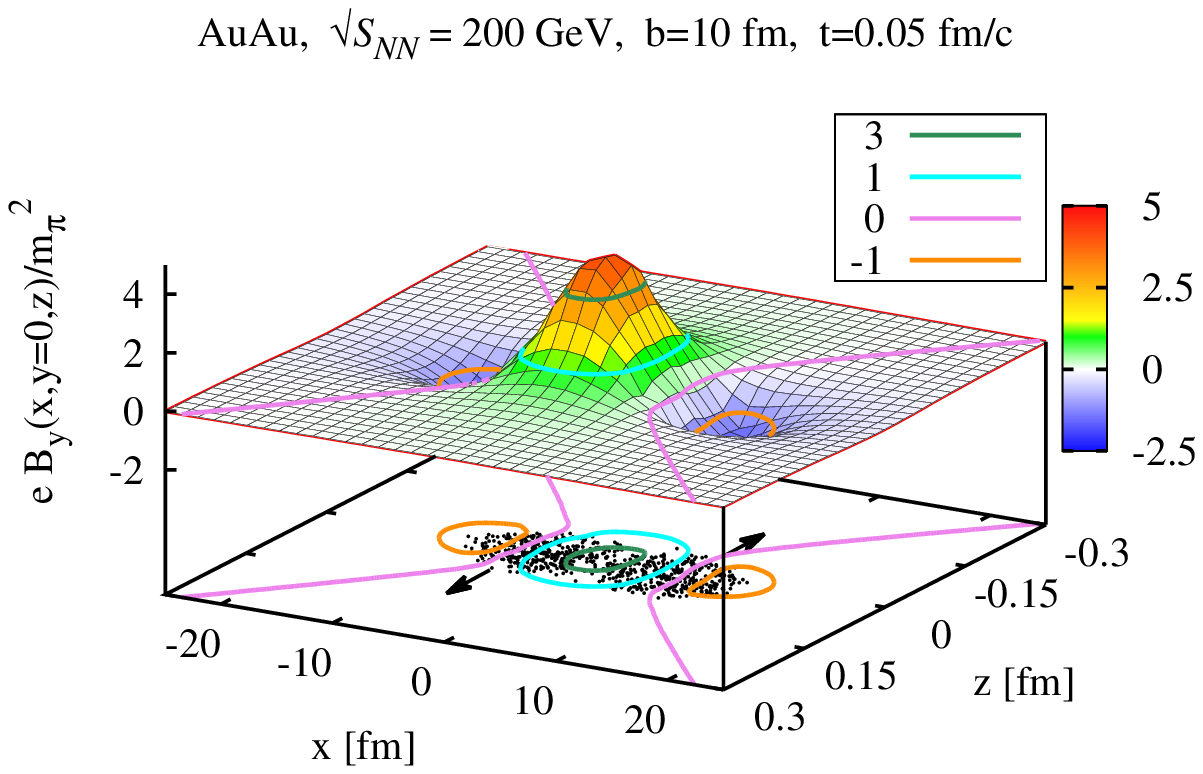}
\includegraphics[height=6.0truecm] {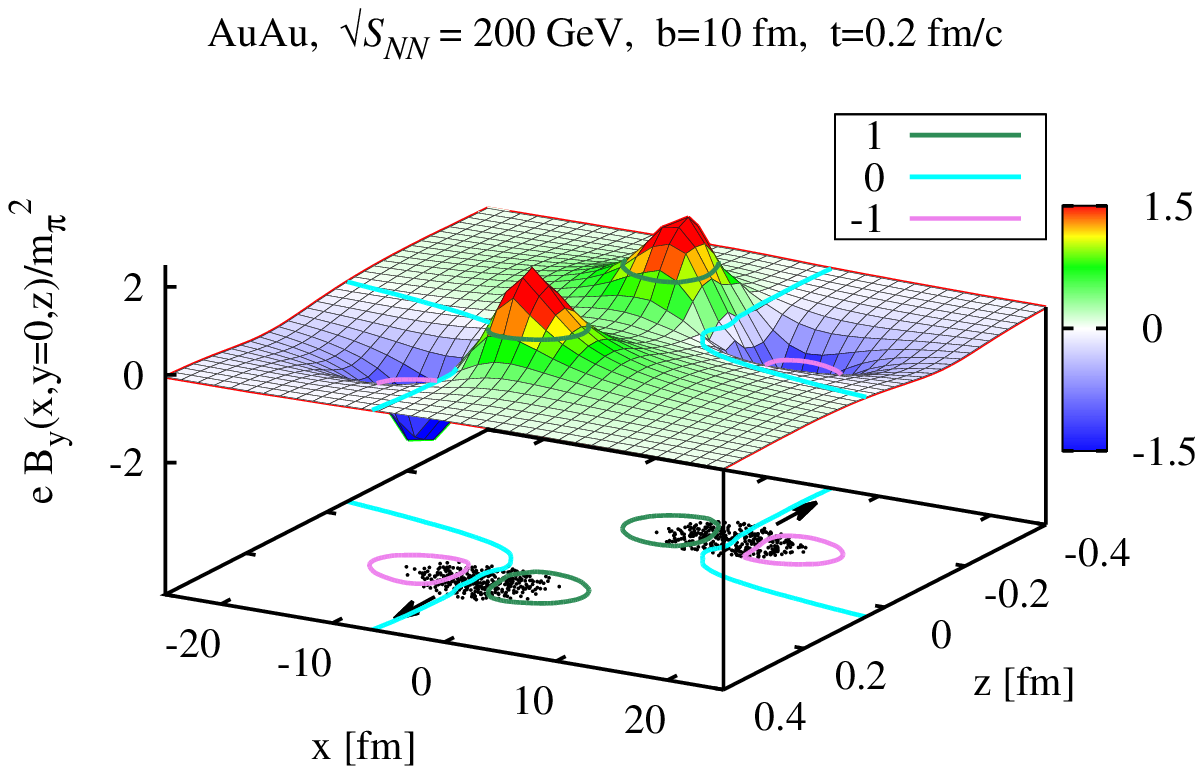}
\hspace{0.1cm}
\includegraphics[height=6.0truecm] {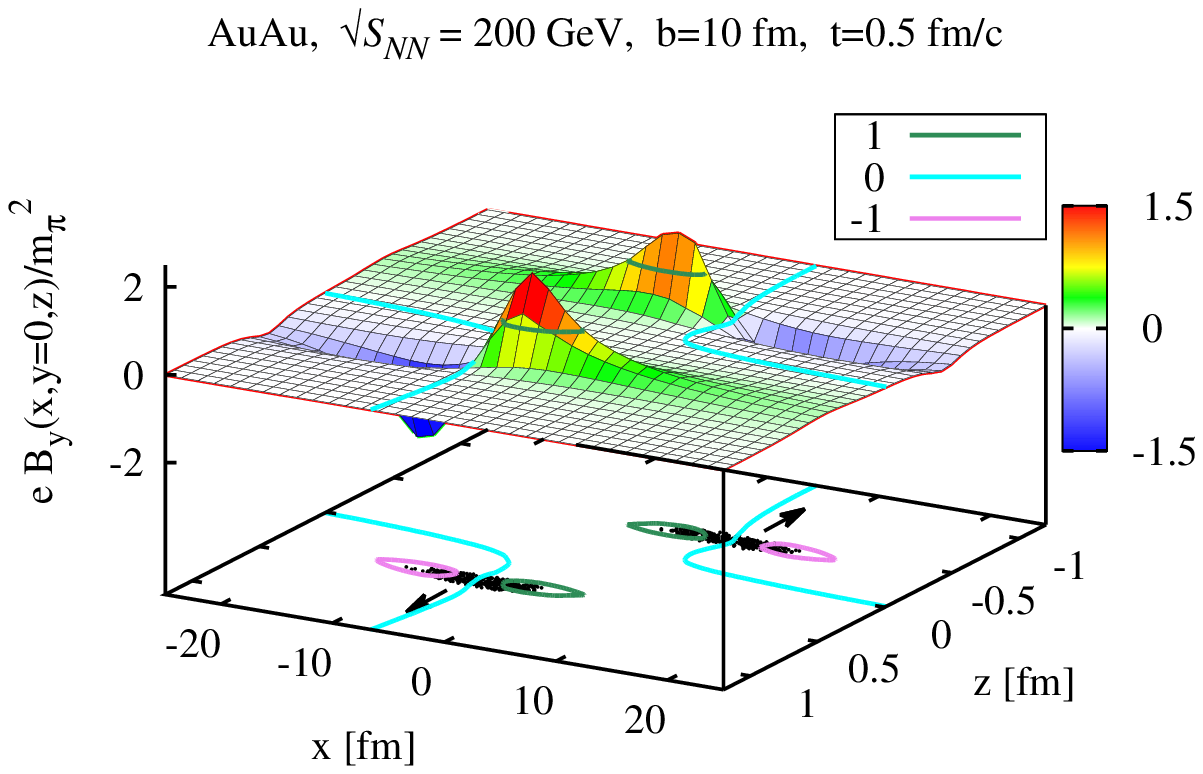}
\caption{(Color online) Time dependence of the spatial distribution of the magnetic
field $B_y$ at times $t$ created in Au+Au ($\sqrt{s}$=200 GeV)
collisions with the impact parameter $b=$10 fm. The location of
spectator protons is shown by dots in the $(x-z)$-plane. The level
$B_y=0$ and the projection of its location on the $(x-z)$ plane are
shown by the solid lines.
 }
\label{By_d}
\end{figure*}

\end{widetext}

 The time evolution of
$eB_y(x,y=0,z)$ for Au+Au collisions for the colliding energy
$\sqrt{s_{NN}}=$200 GeV at the impact parameter $b=$10 fm   is shown
in Fig.~\ref{By_d}. If  the impact parameter direction is taken as
the $x$ axis (as in the present calculations), then the magnetic
field will be directed along the $y$-axis perpendicularly to the
reaction plane $(z-x)$. The geometry of the colliding system at the moment 
considered is demonstrated by points in the $(z-x)$ plain
where every point corresponds to a spectator nucleon. It is seen
that the largest values of $eB_y\sim 5 m_\pi^2$ are reached in the
beginning of  a collision for a very
\noindent  short time corresponding to the
maximal overlap of the colliding ions. Note that this is an
extremely high magnetic field, since $m_\pi^2 \approx 10^{18} {\rm
gauss}$. The first panel in Fig.~\ref{By_d} is taken at a very early
compression stage with $t=$0.01 fm/c. The time $t=$0.05 fm/c is
close to the maximal  overlapping and the magnetic field here is  maximal.
Then, the system expands (note the different $z$-scales in
different panels of Fig.~\ref{By_d}) and the magnetic field
decreases. For $b=$0 the overlapping time is maximal and roughly
given by $2R/\gamma_c$ which for our case is about 0.15 fm/c. For
peripheral collisions this time is even shorter.

Globally, the spatial distribution of the magnetic field is
evidently inhomogeneous and Lorentz-contracted along the $z$-axis.
At the compression stage there is a single maximum which in the
expansion stage is splitted into two parts associated with the
spectators. In the transverse direction the bulk magnetic field is
limited by two minima coming from the torqued structure of the
single-charge field (see Fig.~\ref{oneCh}).

\begin{figure}[thb]
\includegraphics[height=6.0truecm,clip=] {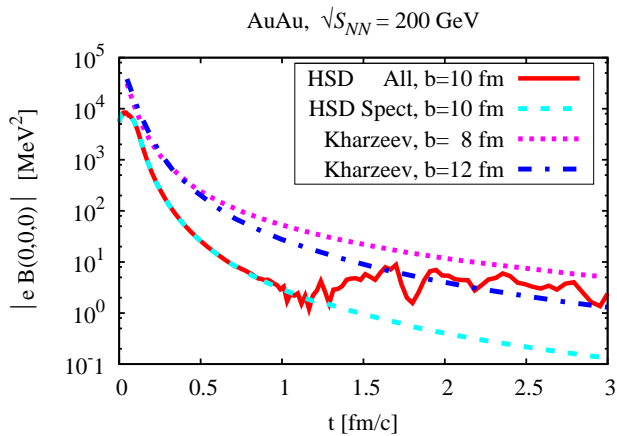}
\caption{(Color online) Time dependence of the $|eB|$ field in the center of the
nuclear overlap region for  Au+Au($\sqrt{s}$=200 GeV) collisions
from the HSD calculations. The dotted and dot-dashed curves are from
Ref.~\cite{KMcLW07} at the impact parameters $b=$8 and 12  fm,
respectively.
 }
\label{E_Kh}
\end{figure}

The possibility of attaining extremely high magnetic fields in
heavy-ion collisions was pointed out 30 years
ago~\cite{VA80} but there have been only two real attempts to
estimate the magnetic field for relativistic heavy-ion
collisions~\cite{KMcLW07,SIT09}. In Ref.~\cite{KMcLW07} the
colliding ions were treated as infinitely thin layers
(pancake-like), and the results in the center of  a Au-Au collision
$eB_y(0,0,z)$ could be presented in a semi-analytical form. In
Fig.~\ref{E_Kh} these estimates are confronted with our results. It
is
clearly seen that the magnetic field in our transport model for
$b=$10 fm is lower than the estimate from Ref.~\cite{KMcLW07} for
both $b=$12 and 8 fm. This difference originates mainly from the fact that
to simulate rapidity degradation of pancake-like nuclei, a heuristic function
was assumed with making no difference between surviving baryons and
new created particles~\cite{KMcLW07} whereas in our case the dynamical 
hadron-string model 
is used for both primary and subsequent interactions while keeping electric and 
baryonic charges and energy-momentum conservation~\cite{HSD}. The approximation 
of Ref.~\cite{KMcLW07} is reasonable for first collisions but gets progressively 
worse with interaction time as seen in Fig.~\ref{E_Kh}.  The difference in the 
shape of the time dependence of the magnetic field for early times is due to
neglecting  the finite size of the colliding nuclei in Ref.~\cite{KMcLW07}. 
\begin{figure}[t]
\includegraphics[height=6.0truecm] {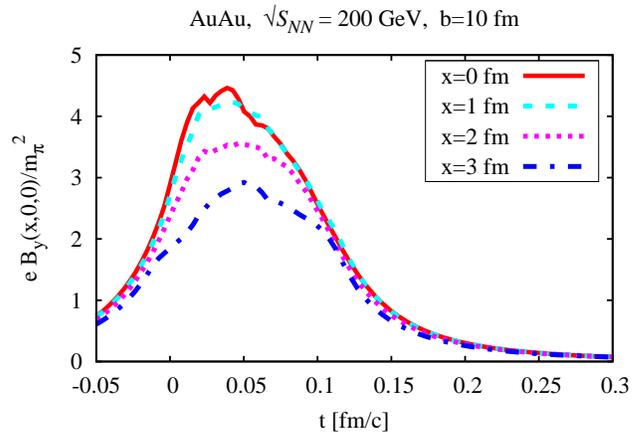}
\caption{(Color online) The magnetic field evolution at the  point $x$ for $y=$0.
 }
\label{Bx-T}
\end{figure}
\begin{figure}[h!]
\includegraphics[height=6.0truecm] {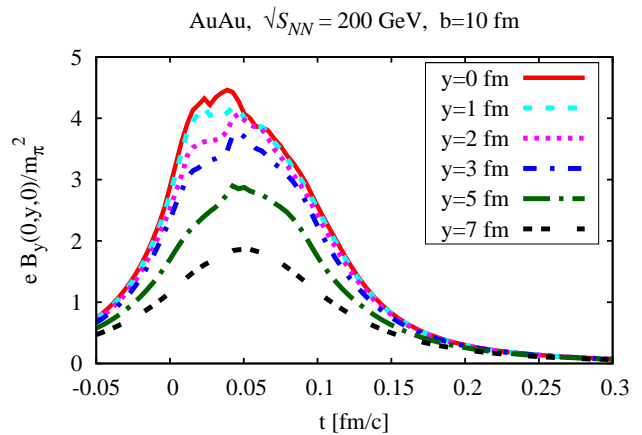}
\caption{(Color online) Time evolution of the magnetic field at the  point $y$ for
the central overlap point $x=$0.
 }
\label{ByYT}
\end{figure}

 Also, in our treatment the self-interaction is excluded for charges within the 
Lorentz-contracted hadron volume. Our consideration treats more accurately  
the retardation effect discussed above which constrains the contributions 
to the given point from some charges. It is especially important for the field 
contribution from participants.

It is of interest to note that in our  transport model, the spectator
contribution to the magnetic field is practically vanishing at
$t\approx $1 fm/c (see Fig.~\ref{E_Kh}). In subsequent times the
magnetic field $eB_y$ is formed essentially due to produced
participants with roughly equal number of negative and positive
charges which approximately compensate each other. The visible
effect in our approach is by an order of magnitude lower than that
in the estimate~\cite{KMcLW07} which demonstrates the essential role
of the retardation in this interaction phase.

 Furthermore, the magnetic field distribution
in~\cite{SIT09} is calculated within the UrQMD model and the
back reaction of the field on particle propagation is disregarded.
Nevertheless, our results are quite close to those of Ref.~\cite{SIT09}.

In Fig.~\ref{Bx-T}, the magnetic field evolution $eB_y(x,y=0,z)$ is
shown as a function of the transverse coordinate $x$. Practically
the difference between results for $x=$0, 1, 2 fm
is less than 20$\%$ except for the boundary of the overlap region
corresponding to $x\approx b/2\sim$5 fm. One thus may conclude that
the magnetic field is rather homogeneous in the transverse
direction.

The magnetic field component $B_y(x=0,y,z)$ along the largest axis
$y$ of the ``almond'' (see Fig.~\ref{tr-pl}) is presented in
Fig.~\ref{ByYT} for different times. The similarity of all curves
for $y\lsim$4 demonstrates a high homogeneity of the created field
$B_y$.
It is of interest that this field stays
almost constant during $\Delta t\sim $0.1 fm/c.

\subsection{Energy density and its correlation with $B_y$}

Along with a high magnetic field, the presence of a quark-gluon
phase is a necessary condition for a manifestation of the chiral 
magnetic effect according to 
\begin{widetext}

\begin{figure*}[thb]
\includegraphics[height=5.truecm] {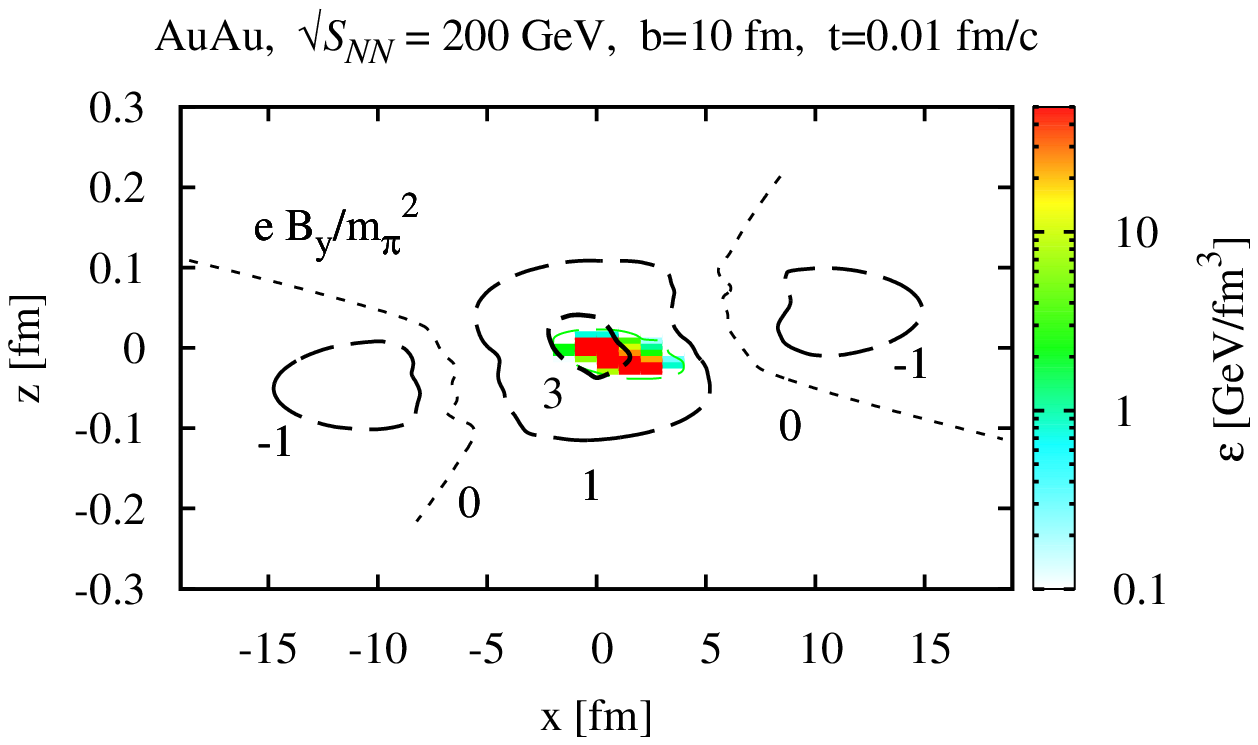}
\hspace{0.1cm}
\includegraphics[height=5.truecm] {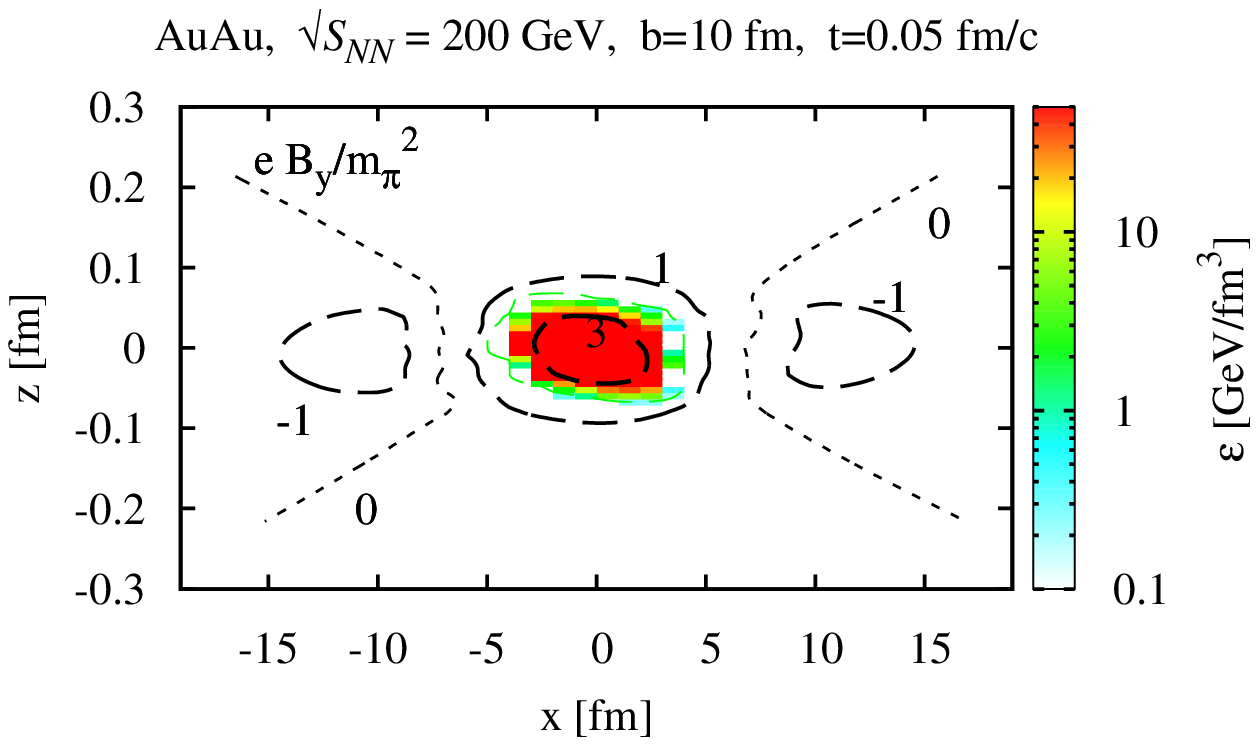}
\includegraphics[height=5.truecm] {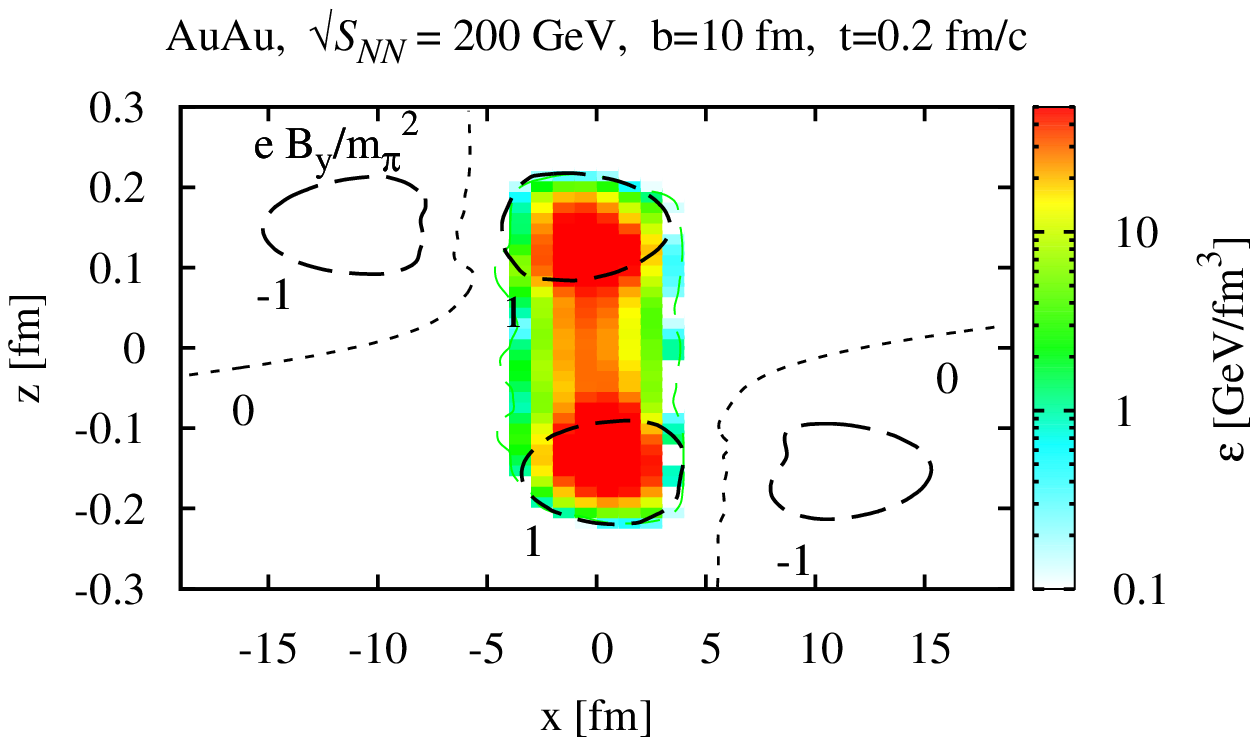}
\hspace{0.1cm}
\includegraphics[height=5.truecm] {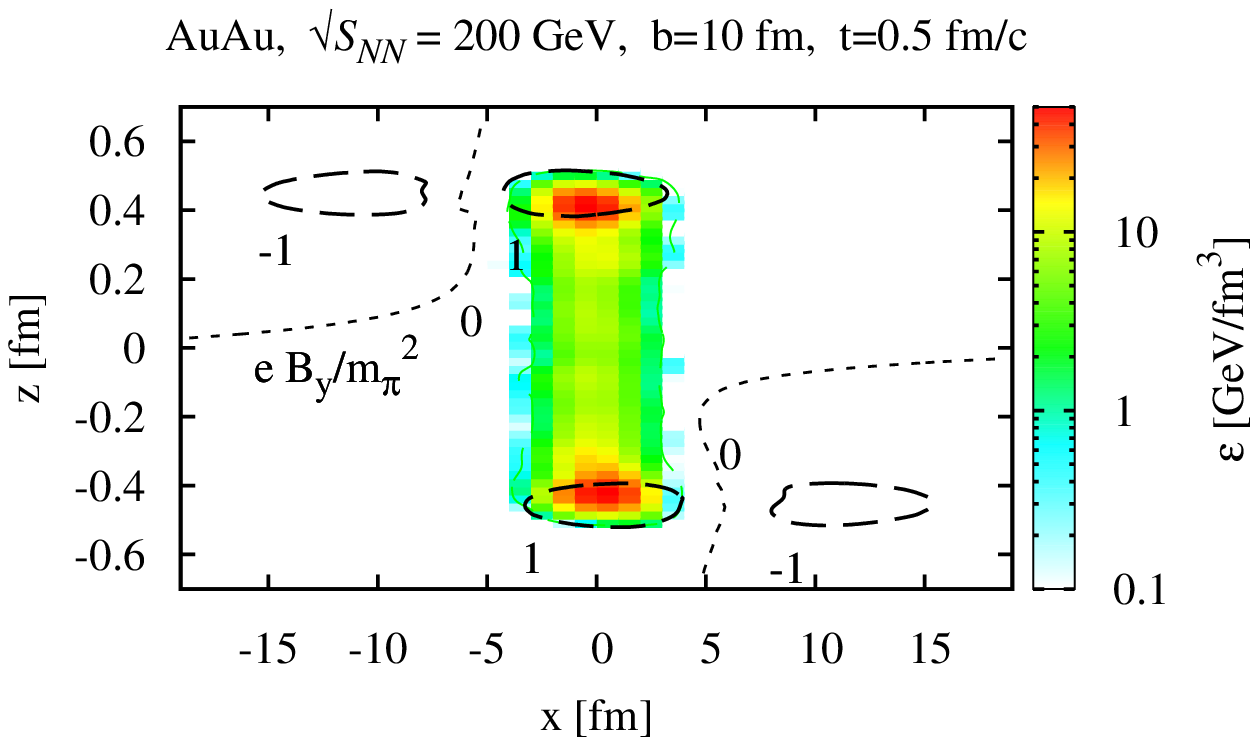}
\caption{(Color online) Spatial correlations in location of the magnetic field and
energy density of participants produced in Au+Au($\sqrt{s}=$200 GeV)
collisions with the impact parameter  $b=$10 fm at the times $t$.
The levels of the magnetic field are plotted by dashed lines whereas
areas with different energy densities are displayed in color.
 }
\label{epsB}
\end{figure*}

\end{widetext}
\noindent   
Refs.~\cite{KZ07,Kharzeev:2004ey,KMcLW07,FKW08,KW09,FRG10}. 
The phase
structure of excited matter is essentially defined by the energy
density (cf. Ref. \cite{PHSD}). One can expect that for energy
densities $\varepsilon \gsim$1 Gev/fm$^3$ the system is in a
deconfined phase. The evolution of the  energy density of created
particles is presented in Fig.~\ref{epsB}. Here the maximal energy 
density (in the center of the colliding system) is $\varepsilon >$
50 GeV/$fm^3$ at the moment of maximal overlap of the nuclei. When
the system expands, it takes a sausage-like shape (or dumb-bell
shaped 
\begin{figure}[h!]
\includegraphics[height=6.0truecm] {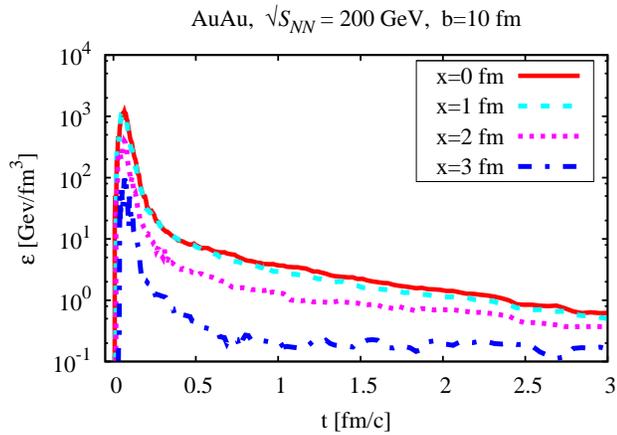}
\caption{(Color online) The average energy density in the Lorentz-contracted cylinder of the
radius $R=$ 1 fm and $|z|<$5$/\gamma$ fm with the $z$-axis passing through
the point $x$.
 }
\label{ExT}
\end{figure}
\noindent if the energy density values are taken into consideration
additionally)   and the energy density drops fast. But even at the
time $t\sim 0.5$ fm/c (last panel in Fig.~\ref{epsB})  the local
energy density  is seen to be above an effective threshold of a
quark-gluon phase transition $\varepsilon\gsim$1 GeV/$fm^3$.
Different levels of the magnetic field strength are plotted in the
same figure. It is clearly seen that the location of the maximum
energy density  correlates with that for the magnetic field.
\begin{figure}[h]
\includegraphics[height=6.0truecm] {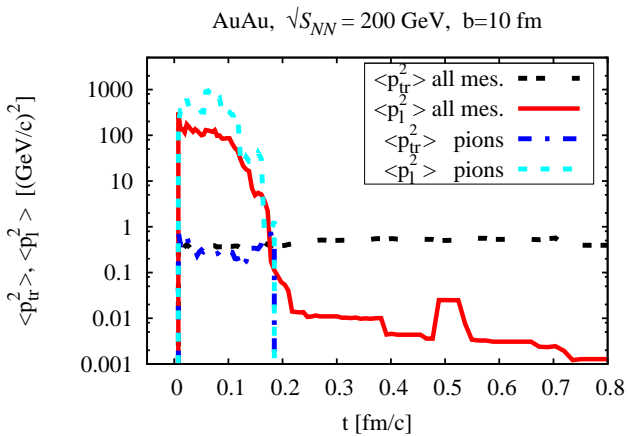}
\caption{(Color online) The average longitudinal and transverse momenta of all mesons and pions
 in the Lorentz-contracted cylinder described above.
 }
\label{Pkv}
\end{figure}

The variation of the energy density distribution with  the
transverse coordinate $x$ is shown in Fig.~\ref{ExT}. Here the
plotted values of $\varepsilon$ correspond to averages within the
Lorentz-contracted cylinder with $|z|<5/\gamma$ fm and radius $R=$1
fm centered at the point $x$. One can see that the energy density
changes more strongly in $x$ than the magnetic field (note the
logarithmic scale in Fig.~\ref{ExT}). In particular, the maximal
$\varepsilon$ decreases by a factor $\sim$20  when one proceeds from
$x=0$ to $x=$3 fm and close to the spectator-participant boundary
(at $x\approx$3 fm) the energy density very quickly (within roughly
$\sim$0.3 fm/c) drops below the effective threshold for
deconfinement $\varepsilon \sim$1 Gev/fm$^3$.

One should note that the energy density should be calculated in the
rest system. The choice of a symmetric position of the cylinder
volume with respect to the $z=$0 plane essentially leads to an
approximately vanishing total momentum of particles inside
this volume. The time  averaged $\gamma$-factor of
particles in this cylinder in the c.m. system is $<\gamma> \sim$1.1.
Note, however, that the created particles are not in local
equilibrium!

In Fig.~\ref{Pkv} the evolution of the average longitudinal $<p_l>$
and transverse $<p_{tr}>$  momentum is shown for mesons in the same
cylinder. All mesons keep a constant transverse momentum $<p_{tr}>$
during the whole evolution even if the fastest pions  escape the
finite cylinder volume by the time $t\sim 0.2$ fm/c. The ratio
$<p_l>/<p_{tr}>$ is very large and does not correspond to that in
equilibrium. The sharp decrease of $<p_l>$ in time is due to fast
mesons streaming-out from the finite volume due to the longitudinal
expansion of the system rather than to equilibration. Thus, in the
early stage $t\lsim$0.2 fm/c the Au+Au system with high energy
density is far from equilibrium. On the one hand, this fact is not
astonishing since this stage should be treated in terms of quarks
and gluons rather than on a hadronic level (cf. corresponding PHSD
studies~\cite{PHSD}). On the other hand, there is a general consensus that local
equilibrium hydrodynamics can be applied to heavy-ion reactions at energies
currently available the BNL Relativistic Heavy Ion Collider (RHIC)  only 
for $t>t_0\approx$0.5 fm/c~\cite{Sh03,KH03}.

\vspace*{2.5cm}
\subsection{Electric field evolution}

\begin{figure*}[thb]
\includegraphics[height=5.50truecm] {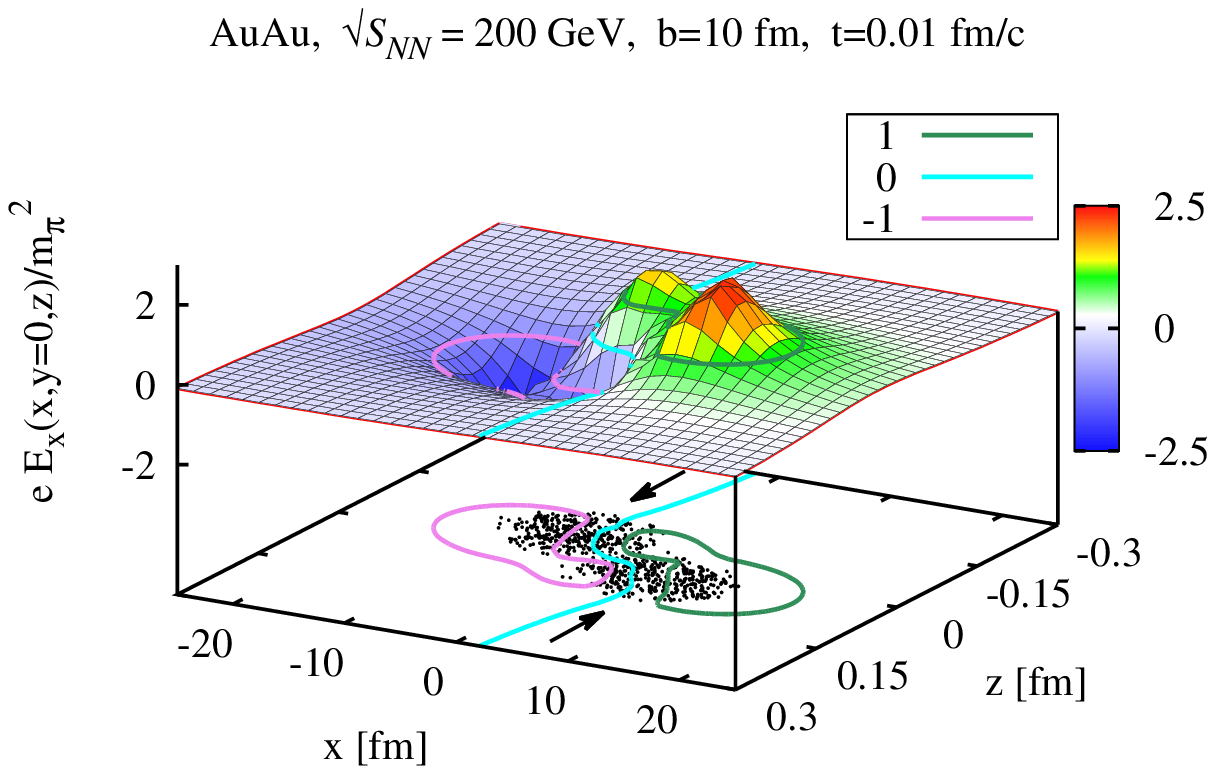}
\hspace{0.1cm}
\includegraphics[height=5.50truecm] {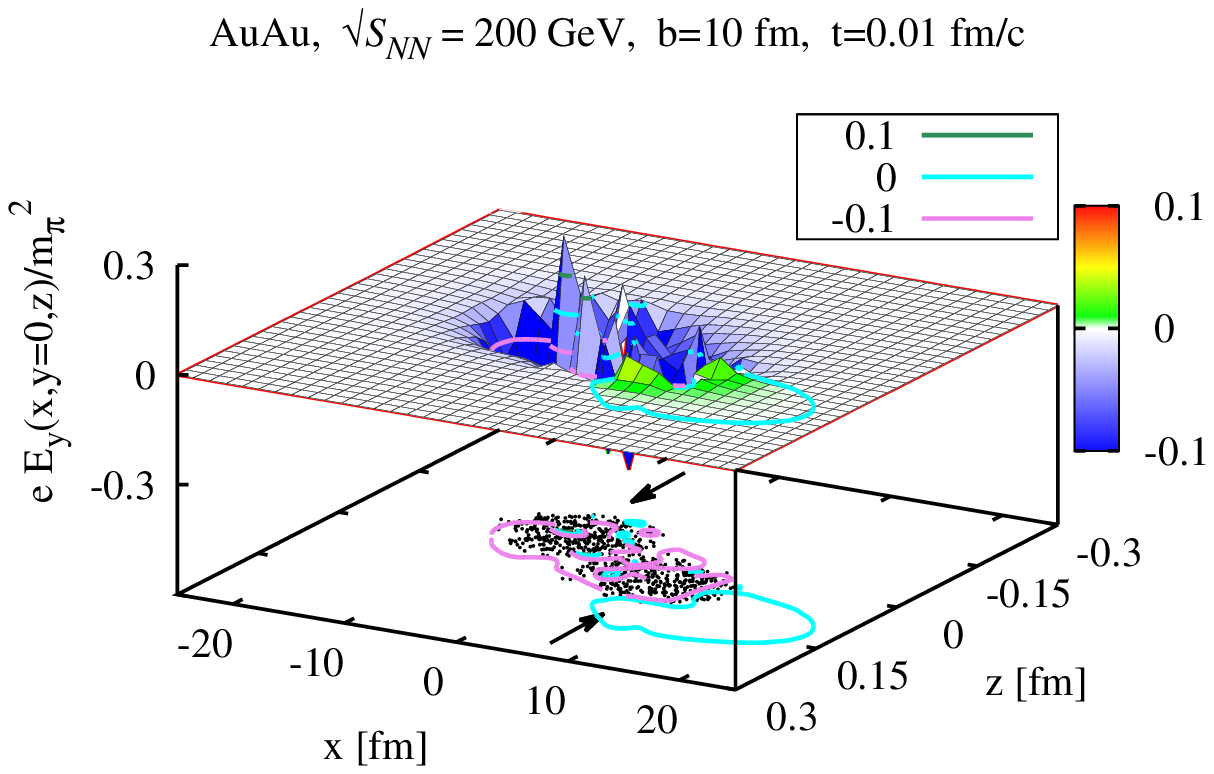}
\includegraphics[height=5.50truecm] {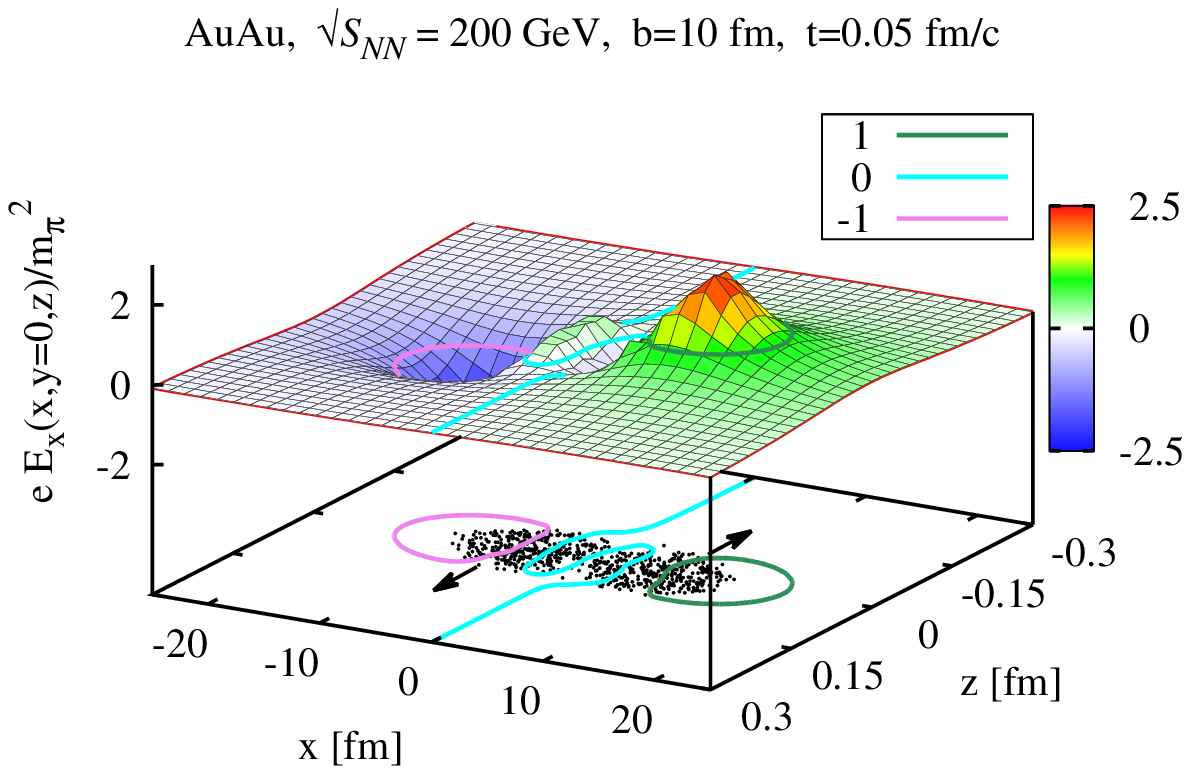}
\hspace{0.1cm}
\includegraphics[height=5.50truecm] {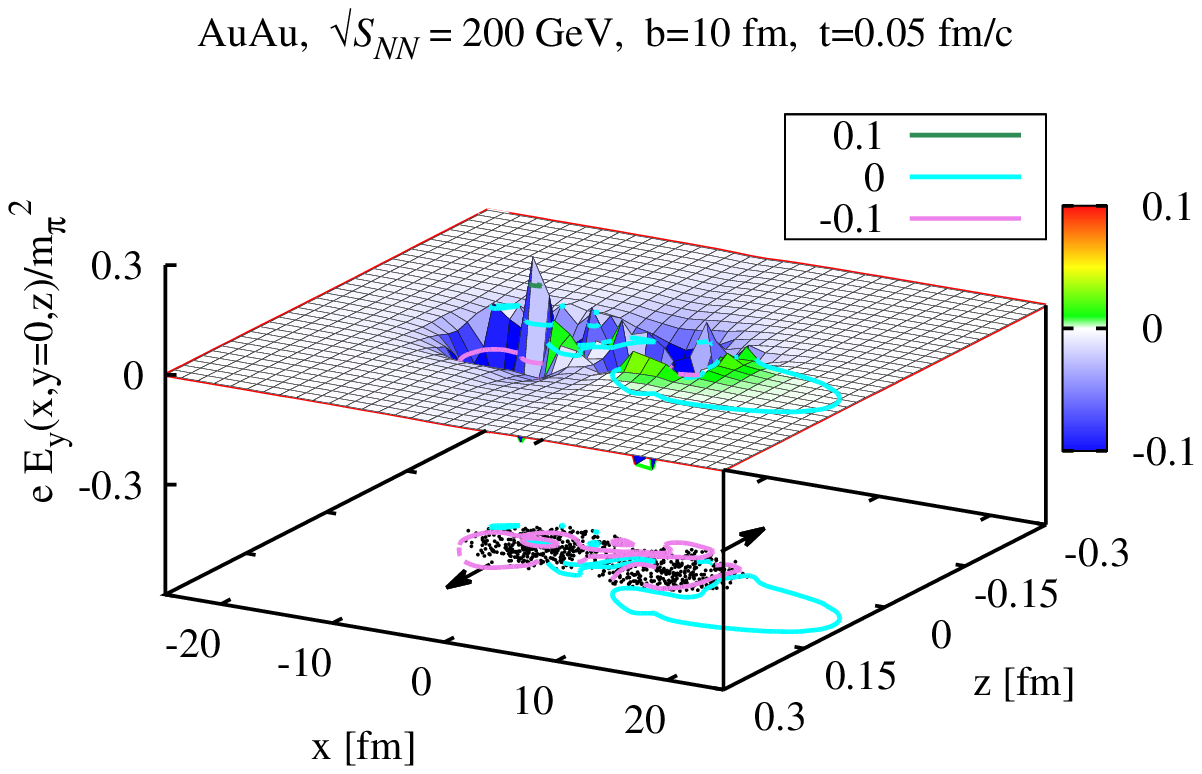}
 \caption{(Color online) Evolution of the $x$- and $y$-components of
the electric field at incoming and maximal overlap in
Au+Au($\sqrt{s}$= 200 GeV) collisions at the impact parameter $b=$10
fm. The $eE_x=$const levels and spectator points are  shown in
the projection on the ($x-z$) plane.
 }
\label{Exy-ev}
\end{figure*}

The background electric field, being orthogonal to the magnetic one,
is directed along the $x$ axis.  The evolution of the $eE_x$ field
for peripheral ($b=$10 fm) collisions of Au+Au at the top RHIC
energy is presented in Fig.~\ref{Exy-ev}. Similar to 
the case of the magnetic field,
the $eE_x(x,y=0,z)$ distribution is also inhomogeneous and closely
correlates with geometry while the field strength looks 
``hedgehog'' shaped. When the two nuclei collide, the electric
fields in the overlap region significantly compensate each other, and
the electric field ${\bf E}$ in the target and projectile 
spectator parts have opposite
\begin{figure*}[thb]
\includegraphics[height=6.0truecm] {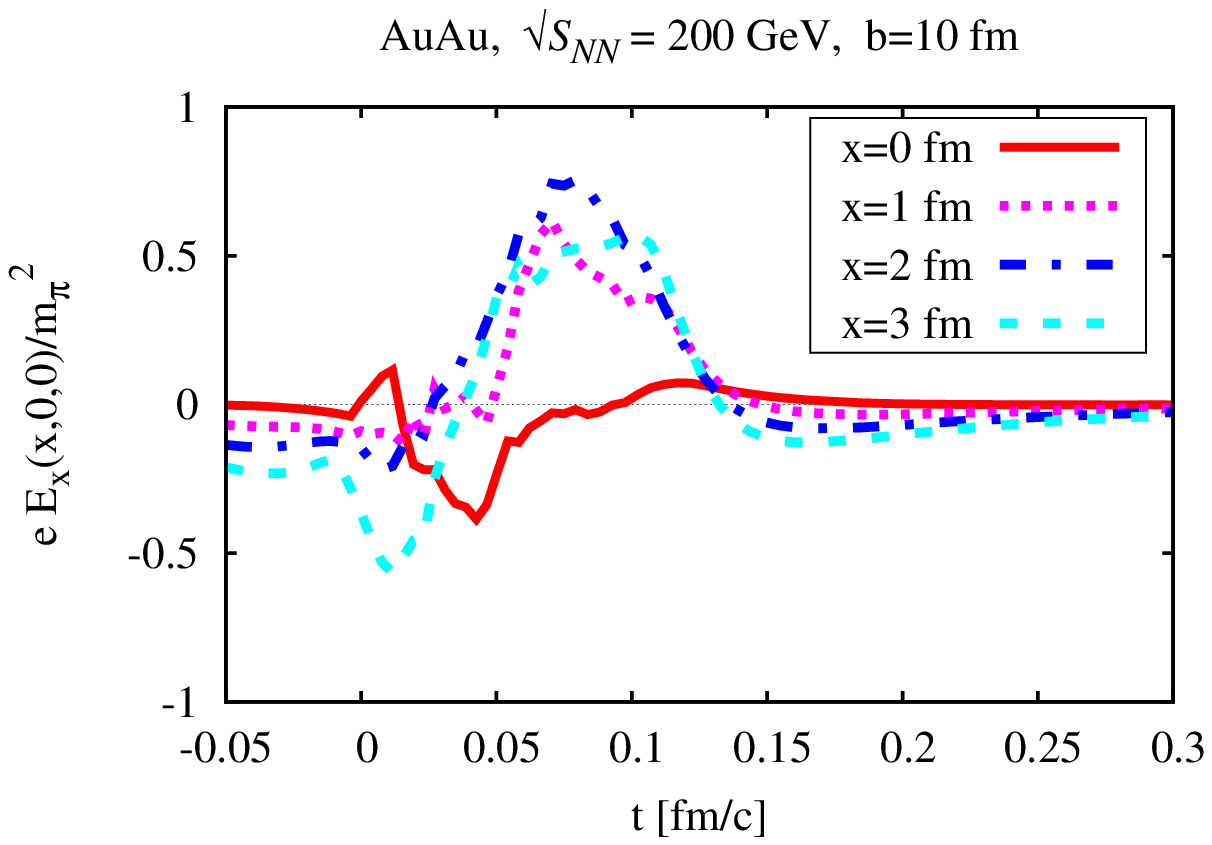}
\hspace{0.1cm}
\includegraphics[height=6.0truecm] {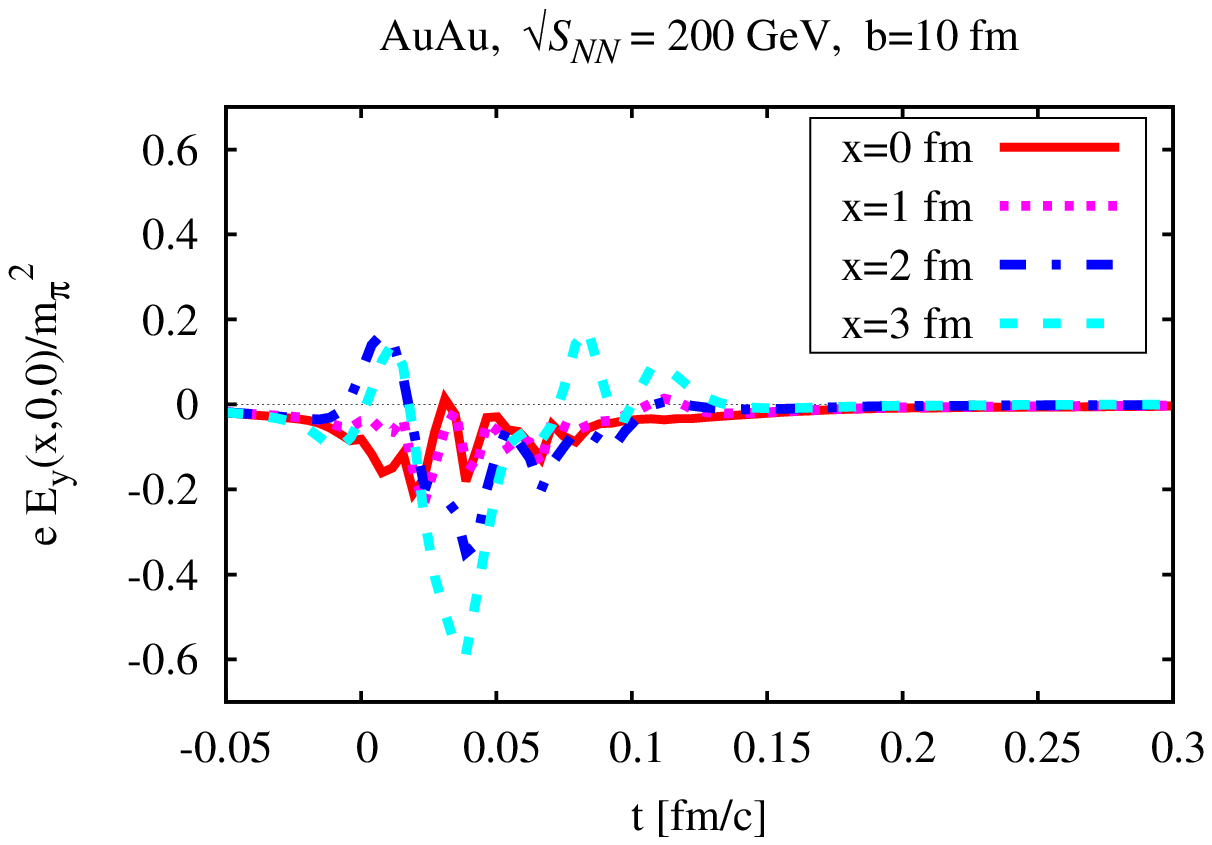}
 \caption{(Color online) Evolution of the $x$ (left panel)- and $y$ 
 (right panel)-components of
the electric field for different values of the transverse coordinate
$x$.
 }
\label{Exy}
\end{figure*}
\noindent signs. As a result, the locations of the maximum/minimum are not in
the central point of the overlap region - as they are for the magnetic field
- but shifted slightly outside. The maximum of the electric field
can be quite large. All these features are seen explicitly in
Fig.~\ref{Exy} where the temporal evolutions of $eE_x(x,0,0)$ and
$eE_y(x,0,0)$ are given for different values of the transverse
coordinate $x$. Due to destructive interference or the ``hedgehog'' 
effect, the electric field
in the central part of the overlap region ($x\approx$0 fm) is
consistent with zero apart from a short period just before reaching
maximal overlap. For $x\simeq$1-3 fm the electric field has a
distinct maximum of  $eE_x/m^2_\pi\approx$(0.5-0.6) which is only by
a factor of about 10 less than the maximal magnetic field $eB_y$
(cf. Fig.~\ref{Bx-T}). But when moving farther away from the center
of the overlap region the $eE_x$ component drops sharply and then
becomes  negative. The $eE_y$ field component is quite small for the
central part of the overlap region and increases slightly for larger
$x$. Note that the electric field is negligible for $t\gsim$0.15
fm/c.


\subsection{Scalar product of $({\bf E}\cdot{\bf B})$ }

\begin{figure*}[thb]
\includegraphics[height=5.20truecm] {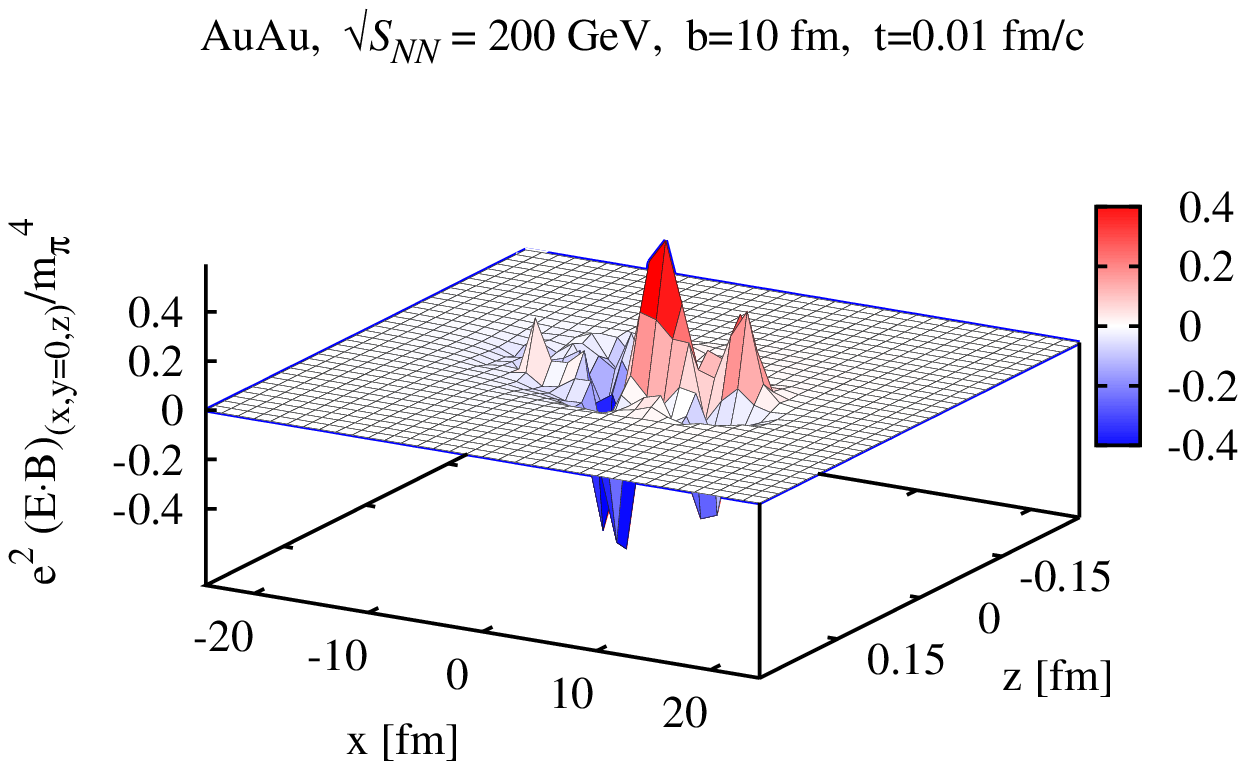}
\hspace{0.1cm}
\includegraphics[height=5.20truecm] {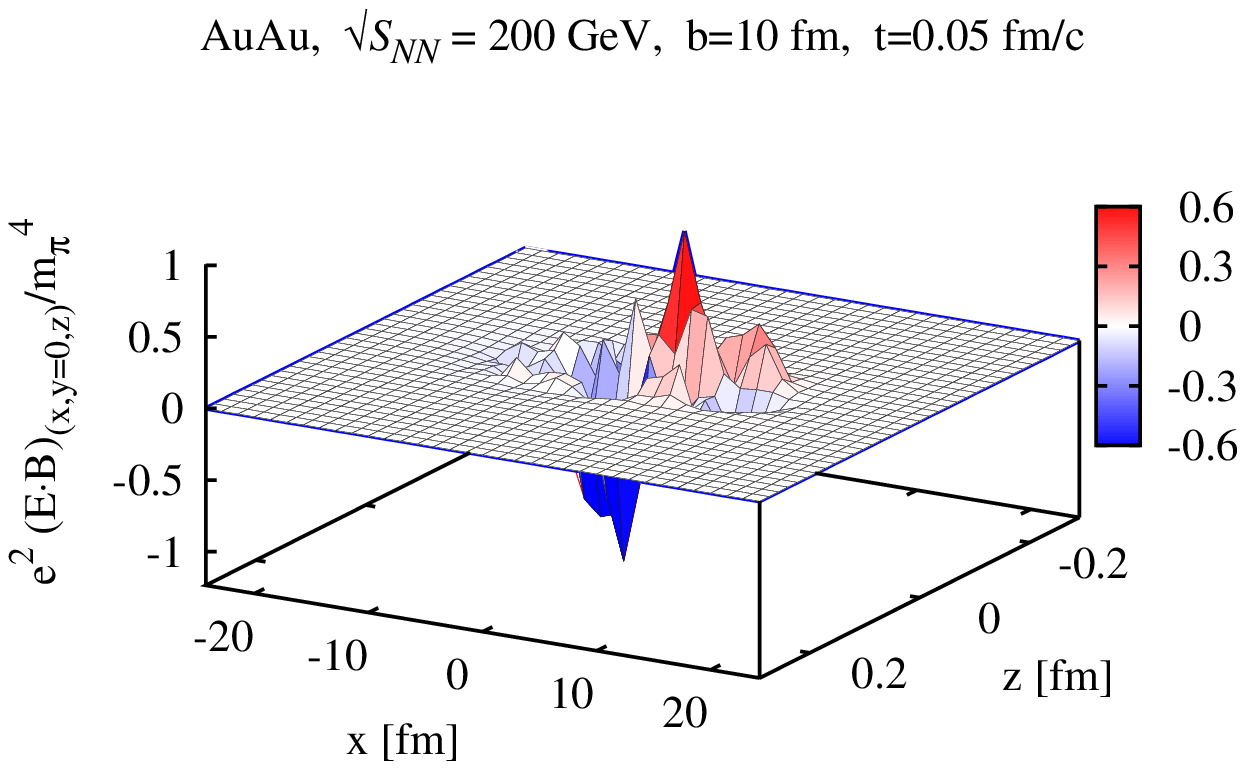}
\includegraphics[height=5.20truecm] {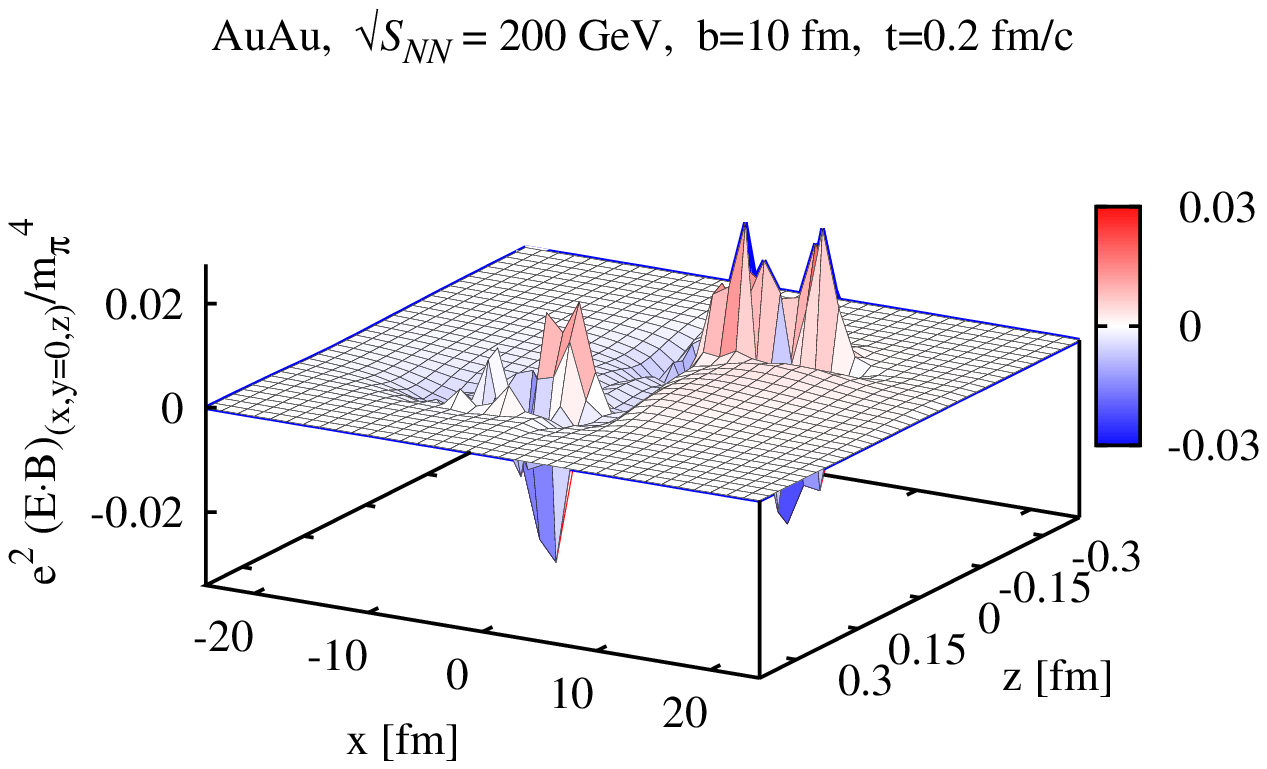}
\hspace{0.1cm}
\includegraphics[height=5.20truecm] {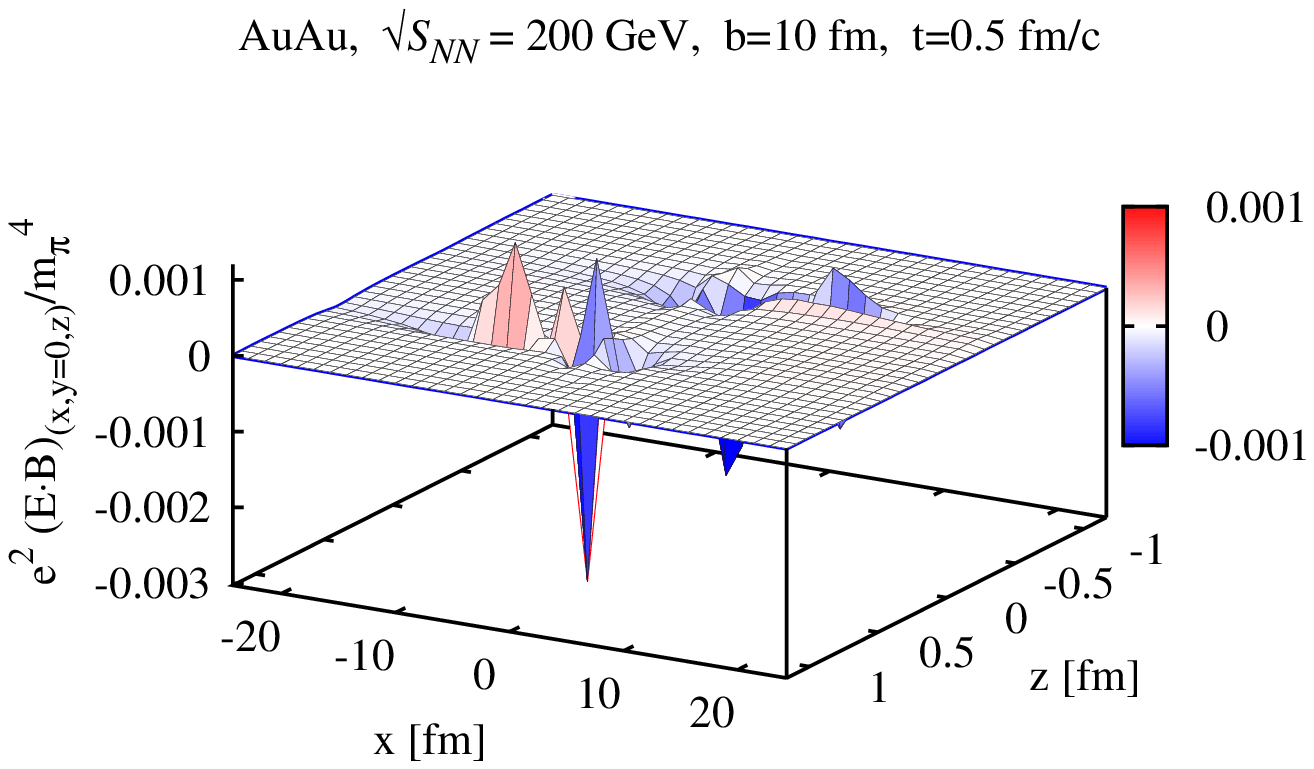}
\caption{(Color online) Space-time evolution of of the scalar product of electric
and magnetic  field  $({\bf E}\cdot {\bf B})$ for a
 Au+Au reaction with the impact parameter $b=$10 fm at $\sqrt{s}$=200 GeV.
 }
\label{EtoB}
\end{figure*}

Since the magnetic field is odd under time reversal (or
equivalently, under the combined charge conjugation and parity
$\cal{CP}$ transformation), the time reversal symmetry of a quantum
system is broken in the presence of an external magnetic field. A
magnetic field ${\bf B}$ can also combine with an electric field
${\bf E}$ to form the Lorentz invariant $({\bf E}\cdot {\bf B})$
which changes the sign under a parity transformation. In the normal
QCD vacuum with its spontaneously
broken chiral symmetry  the leading interaction involves the
invariant $({\bf E}\cdot {\bf B})$ which enters {\it e.g.} into the matrix
element that mediates the two-photon decay of the neutral
pseudoscalar mesons. In the deconfined chirally symmetric phase of
QCD, the leading interaction term is proportional to
$\alpha\alpha_s({\bf E}\cdot {\bf B})({\bf E^a}\cdot {\bf B^a})$,
where ${\bf E^a}$ and $ {\bf B^a}$ denote the chromoelectric
and chromomagnetic fields, respectively, and $\alpha$ and $\alpha_s$
are the electromagnetic and strong QCD couplings. Both interactions
are closely related to the electromagnetic axial anomaly, which in
turn relates the divergence of the isovector axial current to the
pseudoscalar invariant of the electromagnetic field
(see Ref.~\cite{MS10}). The evolution of the electromagnetic invariant
  - produced in Au+Au ($\sqrt{s}$=200 GeV)
collisions at the impact parameter $b=$10 fm - is shown in
Fig.~\ref{EtoB}.

As seen the electromagnetic invariant $({\bf E}\cdot {\bf B})$ is
non-zero only in the initial time $t\lsim$0.5 fm/c where the 
$({\bf E}\cdot {\bf B})$ distribution is quite irregular and its non-zero 
values correlate well with the location of the overlap region.  For
longer times this electromagnetic invariant vanishes as follows from the 
electric field space-time distributions (cf. Figs.~\ref{Exy-ev} and \ref{Exy}). 
Note that the quantities plotted in Fig.~\ref{EtoB} are dimensionless
and the scaling factor $m_\pi^4 \ [GeV^4]$ is quite small. In a 
topological domain the chromoelectric fields $({\bf E}^a\cdot {\bf B}^a)\ne$0. 
But here non-vanishing values of the electromagnetic invariant are due to highly 
inhomogeneous distributions of electromagnetic fields ${\bf E}$ and ${\bf B}$.

\subsection{Impact parameter dependence}

\begin{figure}[h]
\includegraphics[height=6.10truecm] {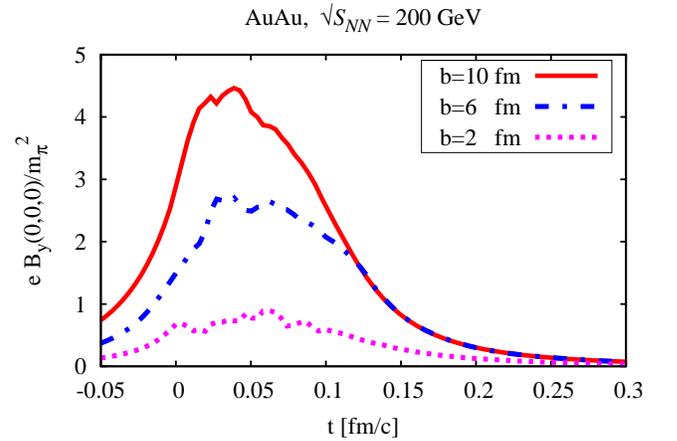}
\caption{(Color online) Impact parameter dependence of the magnetic field in Au+Au collisions at $\sqrt{s_{NN}}=$200 GeV.
 }
\label{bBxT}
\end{figure}

As noted above, the electromagnetic field is formed predominantly by
spectators during the passage time of the two colliding nuclei.
Since the number of spectators increases with the impact parameter
$b$, the magnetic field should also increase for more peripheral
collisions. Indeed, as seen in Fig.~\ref{bBxT}, the magnetic field
decreases gradually with decreasing $b$. When changing the impact
parameter from $b=$10 to 2 fm, the maximal $eB_y(x,0,0)$ decreases
by a factor of 5. For $t\gsim0.3$ fm/c the magnetic field goes down
almost exponentially (cf. with Fig.~\ref{E_Kh}).
\begin{figure}[t]
\includegraphics[width=7.0truecm] {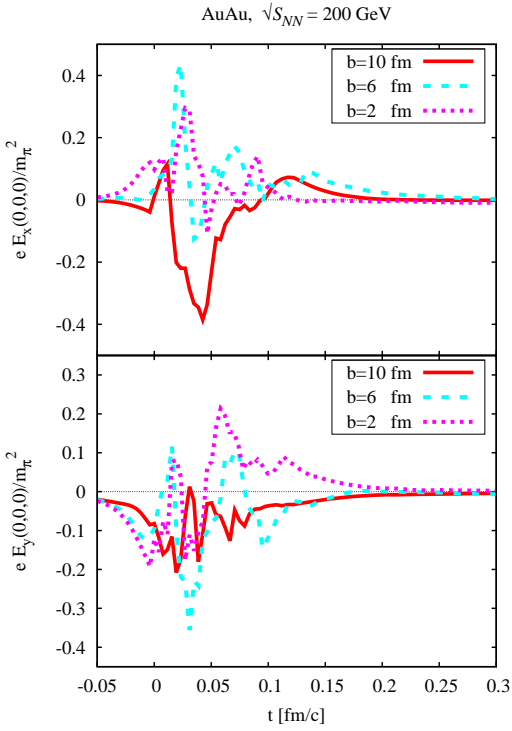}
\caption{(Color online) Impact parameter dependence of the $eE_x(x=0,0,z)$ (top
panel)- and $eE_y(x=0,0,z)$ (bottom panel)- components of the
electric field in Au+Au collisions at $\sqrt{s_{NN}}=$200 GeV.
 }
\label{bExyT}
\end{figure}
\begin{figure}[h!]
\includegraphics[height=6.0truecm] {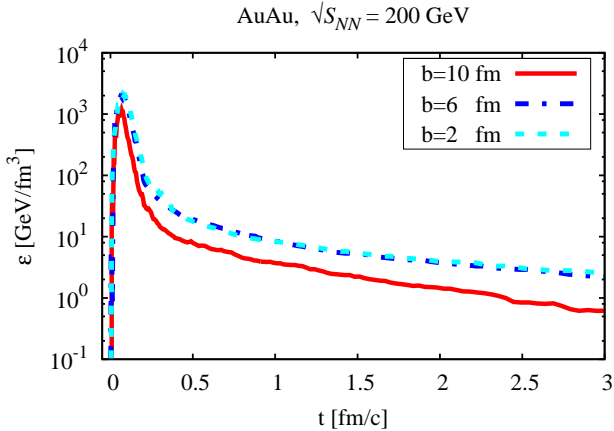}
\caption{(Color online) Impact parameter dependence of the average energy density
in the Lorentz-contracted cylinder of radius $R=$ 1 fm and
$|z|<5/\gamma$ fm with the $z$-axis passing through the point $x=0$ in
 Au+Au collisions at $\sqrt{s_{NN}}=$200 GeV.
 }
\label{bEnxT}
\end{figure}
For the electric field, the $eE_x$-component is only slightly above
the $eE_y$-component;  both are constrained to the time interval
$0\lsim t\lsim$ 0.2 fm/c (Fig.~\ref{bExyT}). Irregularities in these distributions 
are due to the ``hedgehog'' effect mentioned  above.

In contrast, the impact dependence of the energy density
$\varepsilon$ should correlate with the number of participants and
therefore reach a maximum in central collisions. The $b$-dependence
of the $\varepsilon$ temporal distributions is quite weak (see
Fig.~\ref{bEnxT}), and it decreases (by a factor of $\sim$2) only for
 far peripheral collisions, $b\sim$10 fm.

Note that in the $(x-z)$ plane, the location of the maximal values of
the magnetic field and the energy density correlate with each other
(see Fig.~\ref{epsB}).

\subsection{Dependence on collision energy }

In principle, the collision energy dependence of electric and
magnetic fields is given by Eqs. (\ref{E1}) and (\ref{E2}),
respectively, since the fields predominantly emerge from spectators
moving with the initial velocity $v$. These expressions involve
velocity-dependent factors $v$ and $1-v^2$, which  vanish in the
limiting cases $v\to $0 or $v\to$1. In any of these limiting cases,
the denominators stay finite and the energy dependence is given by
the numerator. So, the electric field approaches zero in the
ultrarelativistic limit $v\to$1; the magnetic field vanishes in both
limits  $v\to$0 and $v\to$1.
\begin{figure}[h]
\includegraphics[height=6.0truecm] {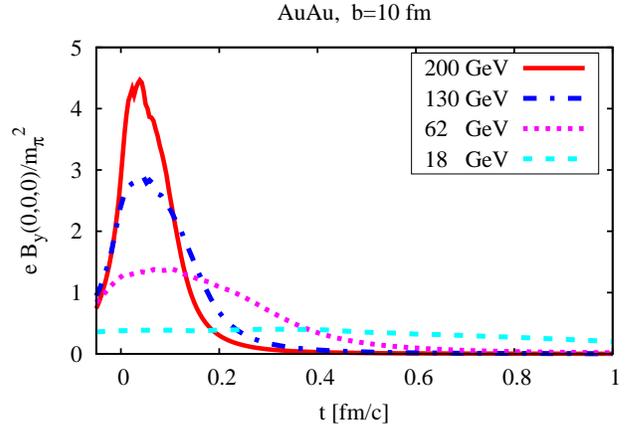}
\caption{(Color online) Collision energy dependence of the magnetic field evolution 
in the center point of the overlap region.
 }
\label{ByTs}
\end{figure}
As follows from Fig.~\ref{ByTs}, the maximal strength of the
magnetic field $eB_y(0,0,0)$ decreases roughly proportionally to
$\sqrt{s_{NN}}$ and at the top energy available at the CERN Super Proton 
Synchrotron (SPS) ($\sqrt{s_{NN}}\approx $18
GeV), it is only about $\sim 0.4/m_\pi^2$ which appears to be too low
to search for the Chiral Magnetic Effect~\cite{TV10}. One should keep
\begin{figure}[h]
\includegraphics[width=8.0truecm] {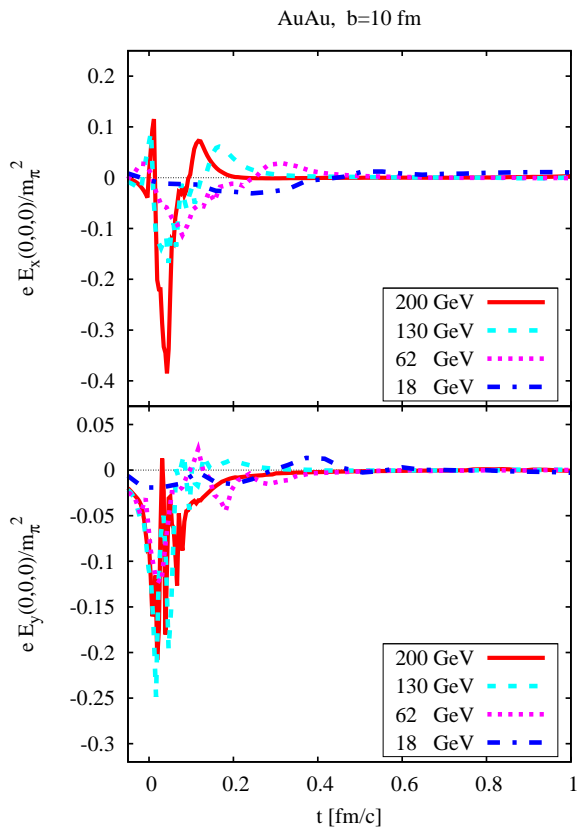}
\caption{(Color online) Collision energy dependence of the $x$ and $y$ component of
the electric field.
 }
\label{ExyTs}
\end{figure}
\begin{figure}[h!]
\includegraphics[height=6.0truecm] {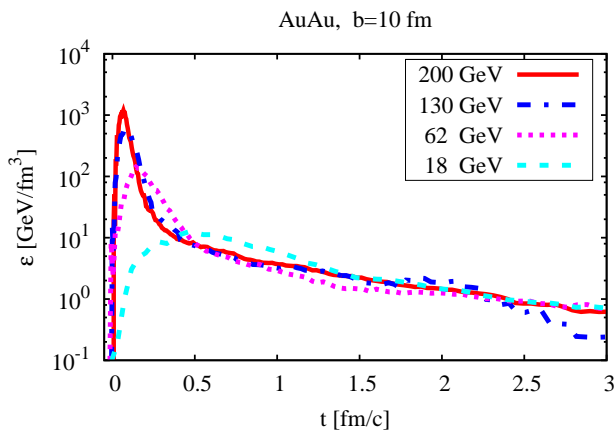}
\caption{(Color online) Collision energy dependence of the average energy density 
in the Lorentz-contracted cylinder of
radius $R=$ 1 fm and $|z|<5/\gamma$ fm with the $z$-axis passing through
the point $x=0$.
 }
\label{EnTs}
\end{figure}
in mind that the chiral magnetic effect  depends not only on
the magnitude of the magnetic field but also on the time of the
system within the high magnetic field.  As  seen in Fig.~\ref{ByTs},
with decreasing collision energy the width of the time distribution
of $B_y$ becomes wider  because it is defined essentially by the
Lorentz-contracted overlap region $\sim \Delta r/ \gamma = 2 m_N
\Delta r /\sqrt{s_{NN}}$. The RHIC energy is not high enough to
see the limiting case corresponding to $(1-v^2)\to$0.

According to our expectation, the maximal strength of the
$eE_x(0,0,0)$ and $eE_y(0,0,0)$ components decrease for lower
collision energy as demonstrated in Fig.~\ref{ExyTs}. It is
clearly seen by comparing the results for $\sqrt{s_{NN}}=$200 and 18
GeV. Note that the $eE_y$ component is directed opposite to the
direction of the magnetic field $eB_y$ and will act against ``the
electric charge separation effect''~\cite{Vol05,KZ07,KMcLW07}.

The maximal energy density drops down by two orders of magnitude when going down
from the RHIC to the SPS energy (see Fig.~\ref{EnTs}). During the
time interval 0.5$\lsim t \lsim$2.5  the energy density $\varepsilon(t)$ evolution
for all energies practically coincide  changing from $\varepsilon\sim$10 to
$\sim$1 GeV/fm$^3$.

\section{Observables and electric charge separation}

The HSD model quite successfully describes many observables in a large range
of the collision energy~\cite{HSD,HSDobserv}. Here we investigate to what extent the
electromagnetic field - incorporated in the HSD approach -  will affect some
observables. We shall limit ourselves to Au+Au collisions at $\sqrt{s_{NN}}=$200
GeV and impact parameter $b=$10 fm. In this case we are not able to restrict
our calculations to times $t<$3 fm/c as above but have to calculate the whole
nuclear interaction including the decays of resonances at least up to times 
of 50 fm/c.

\begin{figure}[th]
\includegraphics[height=6.0truecm] {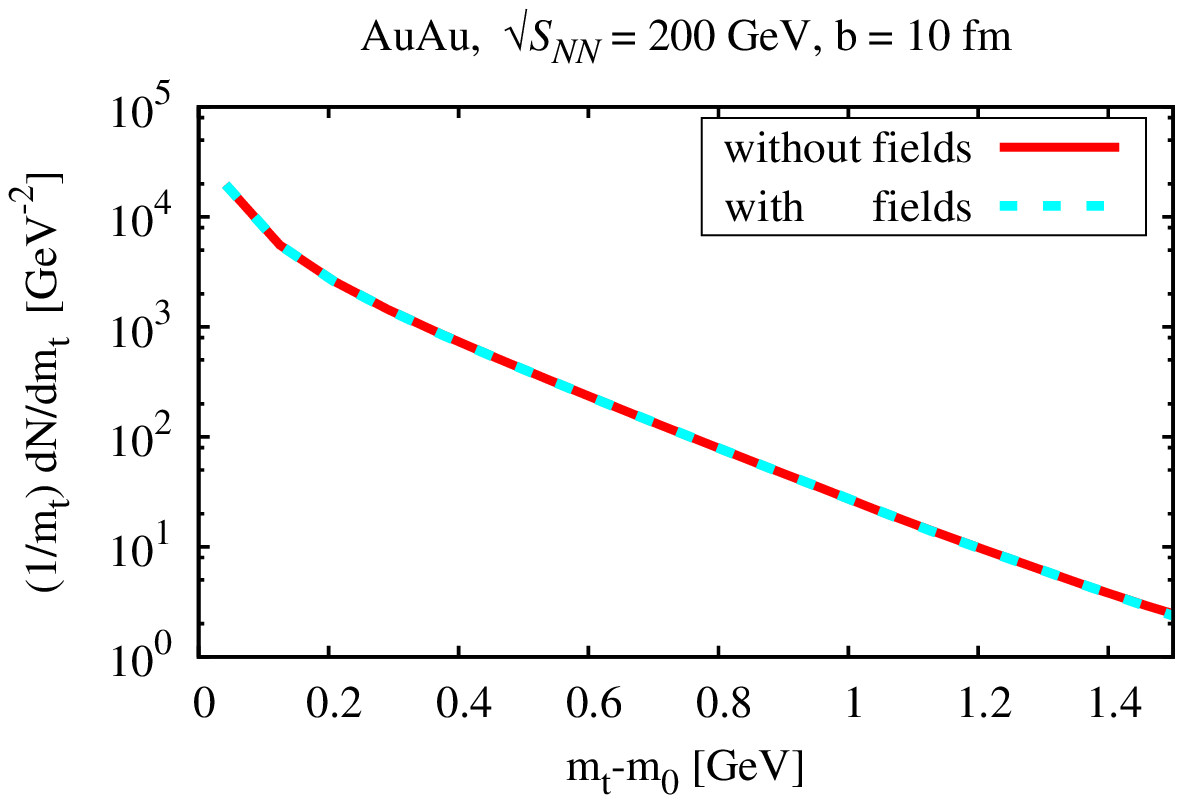}
\includegraphics[height=6.0truecm] {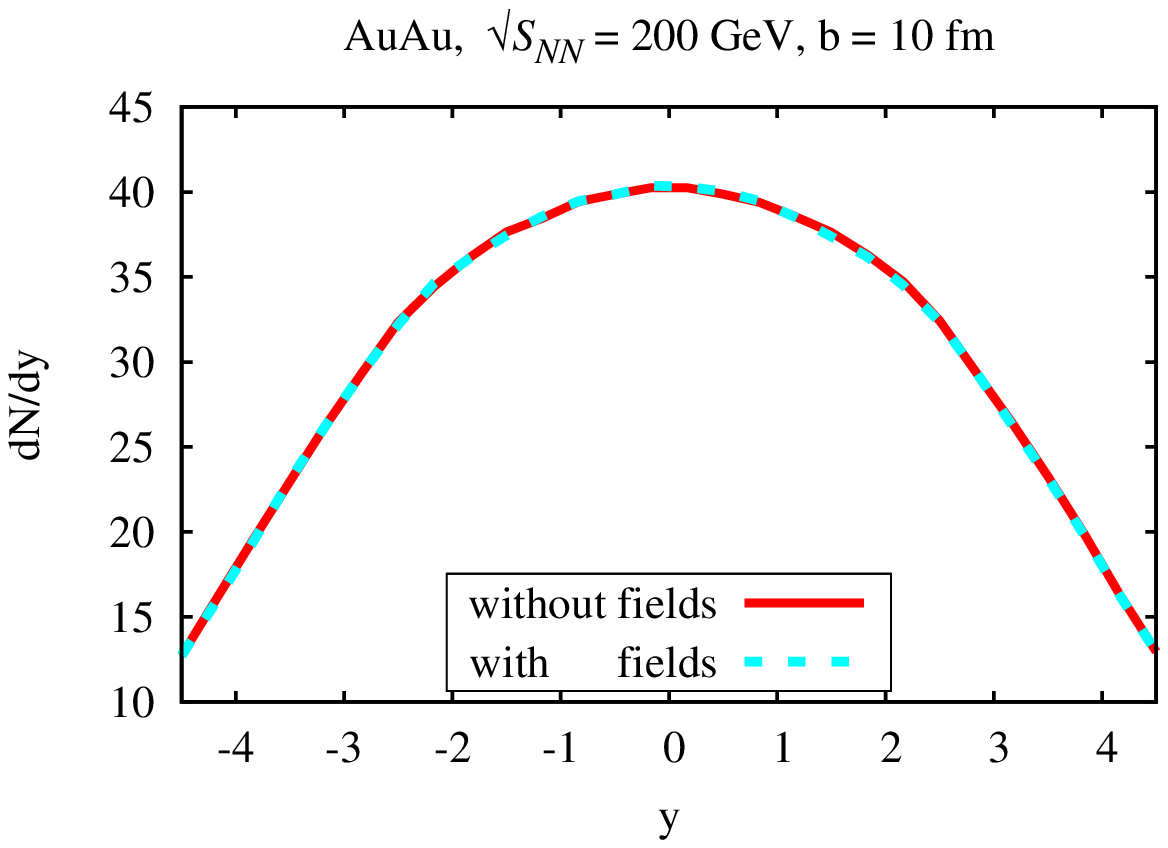}
 \caption{(Color online) Transverse mass and rapidity distributions of charged pions
produced in Au+Au ($\sqrt{s_{NN}}=$200 GeV) collisions at $b=$10 fm. The 
results calculated with and without electromagnetic field are plotted by the  
dotted and solid lines, respectively.
 }
\label{MtY}
\end{figure}

The HSD results for the versions without  and with electromagnetic field - taking
into account the back reaction of the electrodynamic field on the particle 
propagation - are presented in Fig.~\ref{MtY}. With a high degree of accuracy,
\noindent  we see no difference between these
two versions in the transverse mass $m_t$ and rapidity $y$.

\begin{figure}[h!]
\includegraphics[height=6.truecm] {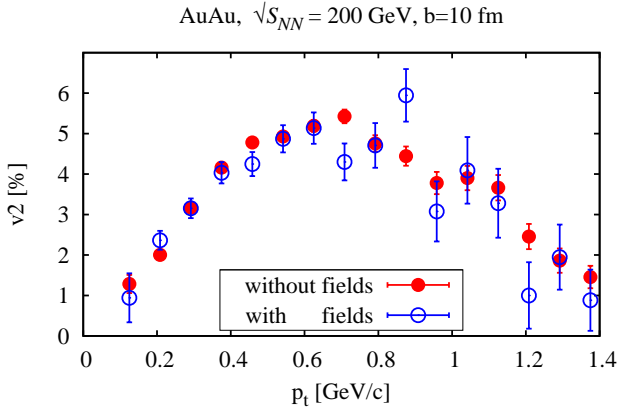}
\caption{(Color online) The transverse momentum dependence of the elliptic flow 
for Au+Au ($\sqrt{s_{NN}}=$200 GeV) collisions at $b=$10 fm.
 }
\label{v2pt}
\end{figure}

In Fig.~\ref{v2pt} the transverse momentum dependence of the elliptic flow
of charged pions is compared for two versions (with and without field) of the 
HSD model. We do not observe any significant difference between the two cases. 
Slight differences are 
seen in the range of $p_{t}\sim$1 GeV/$c$ but certainly it can not be considered as
significant. Note that generally the HSD model underestimates the elliptic flow,
but an inclusion of partonic degrees of freedom within the PHSD approach allows 
it to describe perfectly well the $p_t$ dependence of $v_2$ at the top RHIC 
energy~\cite{PHSD}.  

\begin{figure}[thb]
\includegraphics[height=12.truecm] {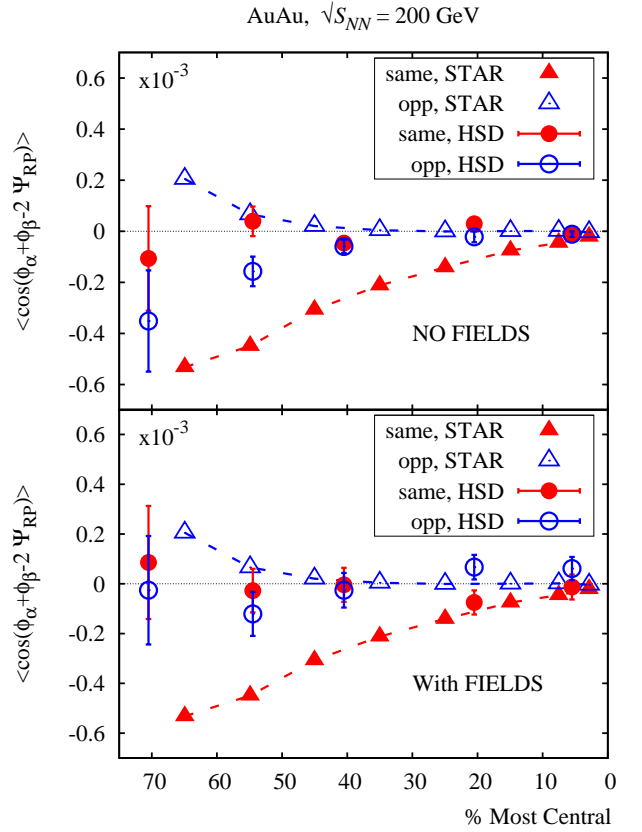}
\caption{(Color online) Azimuthal correlation in the transverse plane as a 
function of centrality
for like and unlike charged pions from Au+Au ($\sqrt{s_{NN}}=$200 GeV) collisions. 
The experimental points - connected by lines - are taken from~\cite{Vol09,STAR-CME}.
 }
\label{C2}
\end{figure}

As a signal of possible $\cal CP$ violations in relativistic heavy-ion collisions,
it was proposed in Ref.~\cite{Vol05} to measure the  two-particle angular
correlation
\begin{equation}
\label{cos}
\langle \cos (\phi_\alpha+\phi_\beta-2\Psi_{RP}) \rangle
\end{equation}
where $\Psi_{RP}$ is the azimuthal angle of the reaction plane defined by
the beam axis and the line joining the centers of colliding nuclei (see
Fig.\ref{tr-pl}). The correlator (\ref{cos}) is calculated on the event-by-event
basis with subsequent  averaging over the whole event ensemble.
The experimental data from the STAR Collaboration~\cite{Vol09,STAR-CME} 
and the results of HSD calculations are presented in Fig.~\ref{C2}. 
The experimental acceptance $|\eta |<1$ and $0.15<p_t<2$ GeV have been also  
incorporated in theoretical calculations.  Note that  the theoretical 
reaction plane is fixed exactly by the initial conditions and therefore is not 
defined by a correlation with a third charged particle as in the 
experiment~\cite{Vol09,STAR-CME}.  The error bars plotted in Fig.\ref{C2} 
show the statistical errors.
The number of events evaluated
with (without) field for the most crucial centralities 0.7 and 0.55 are 
6.8$\cdot 10^4$ and 2.2$\cdot 10^4$  (8.4$\cdot 10^4$ and 5.4$\cdot 10^4$) for 
same charge pion pairs. The computational time for one event with accounting 
for the electromagnetic field is by a factor of about 30  longer than that 
for the case without field.

The expected CME stems from the interplay of topological effects of the excited 
vacuum and the chiral anomaly in the presence of a strong magnetic
field~\cite{KZ07,KMcLW07,FKW08,KW09,FRG10}. One can see that the calculated
background - taking into account hadron string interaction dynamics and evolution 
of the electromagnetic field - is not able to describe the measured distribution 
especially for pions of the  same charge. One should  mention that our
results are rather close to the background estimates within the UrQMD model
in the experimental works~\cite{Vol09,STAR-CME}.


\begin{figure*}[thb]
\includegraphics[width=8.truecm] {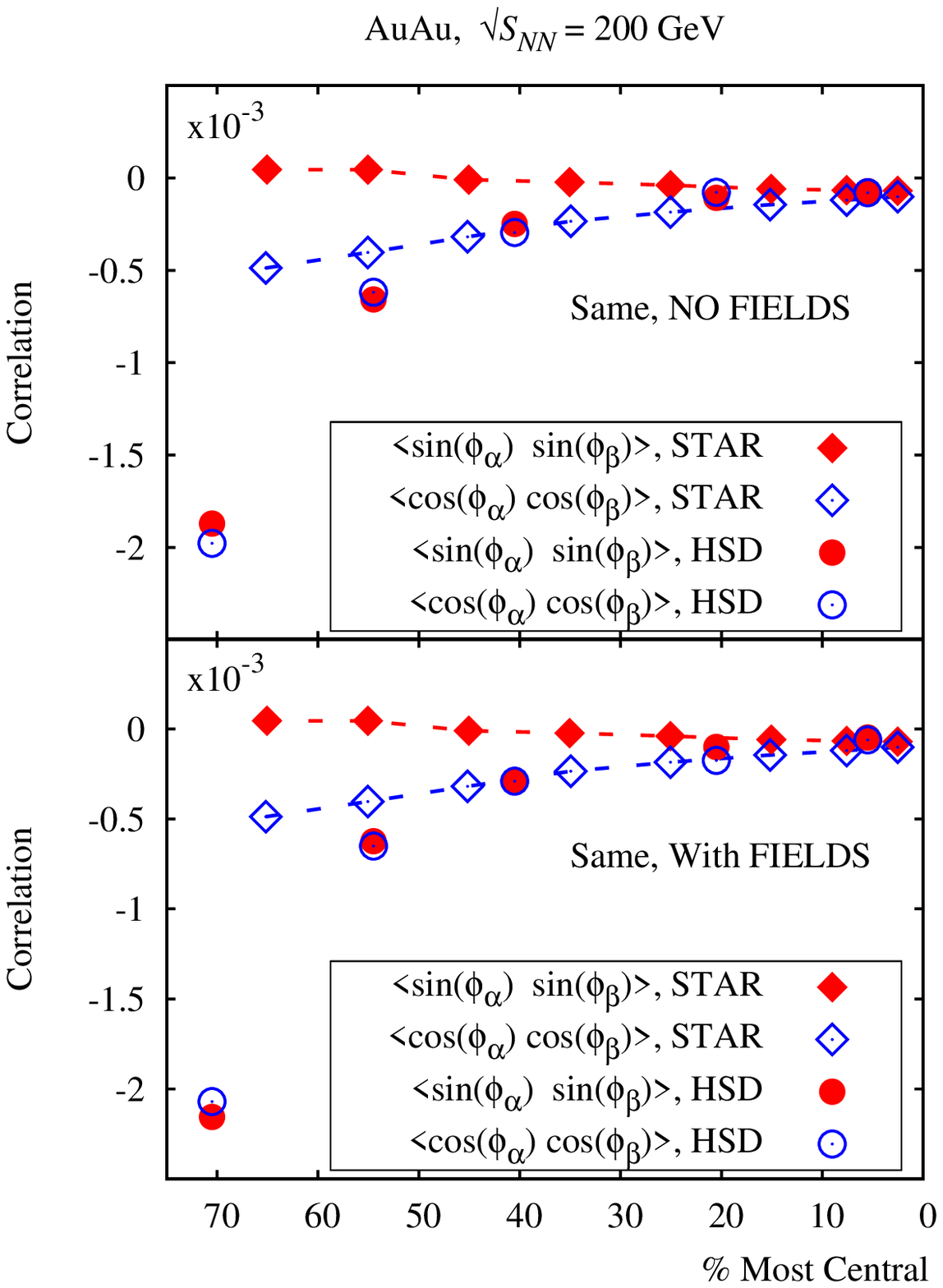}
\includegraphics[width=8.truecm] {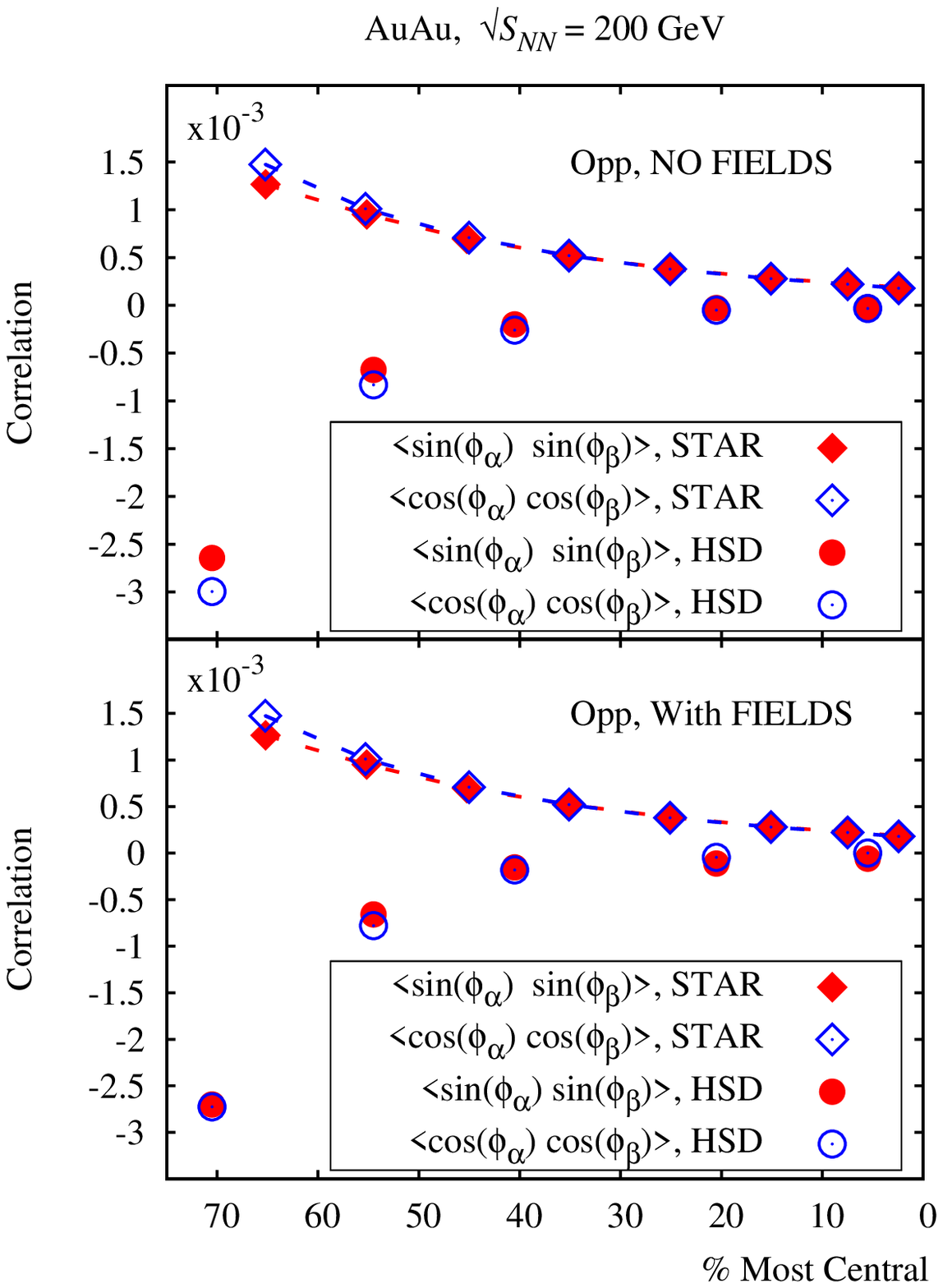}
\caption{(Color online) Projections of azimuthal correlations on the in- and 
out-reaction plane as a function of centrality
for like and unlike charge pions from Au+Au ($\sqrt{s_{NN}}=$200 GeV) collisions. 
The experimental points - connected by lines - are from Ref.~\cite{STAR-CME}.
 }
\label{same-opp}
\end{figure*}

The two particle correlation (\ref{cos}) can be decomposed in ``in-plane'' and
``out-of-plane'' components \footnote{For brevity, below we shall suppress 
$\Psi$ in Eq.(\ref{cos}) but the
azimuthal angle $\phi$ should be measured with respect to the reaction plane.}
\be
\label{cossin}
\langle \cos (\phi_\alpha+\phi_\beta) \rangle &=&
\langle \cos (\phi_\alpha) \cos(\phi_\beta) \rangle \nonumber \\ &-&
\langle \sin (\phi_\alpha) \sin(\phi_\beta) \rangle~.
\ee

Following Ref.~\cite{BKL09} in Fig.~\ref{same-opp} these components are 
 presented for the same $(+,+), (-,-)$ and opposite $(+,-)$ charged pion 
 pairs.  First, there is no difference for HSD 
results without field (top panels in Fig.~\ref{same-opp})and with the 
electromagnetic field (bottom panels). To be more specific, we will 
discuss below the results with the electromagnetic field included. Second, 
the calculated difference between the in-plane ($cosine$-term) and out-of-plane
($sine$-term) components is small in the same and opposite charge  cases.

 Since the
observed correlation (\ref{cossin}) is the difference of these two terms, the
calculated correlation is small as well. Furthermore, for the same charge pairs the
measured $sine$ term is essentially zero while the $cosine$ term is finite. This
implies that the observed correlations are in-plane rather than out-of-plane, as
expected. It is of interest that the measured and calculated $cosine$ terms  
coincide with each other for centralities $\lsim$0.55.
As was noted in Ref.~\cite{BKL09}, the zero $sine$ component is 
contrary to the expectation from the CME, which for the same charge correlation 
results in an out-of-plane correlation. In the
HSD model the $sine$ term is not zero but negative. This is not a surprise
because the induced chromoelectric field parallel to the out-of-plane ${B_y}$
is not included into our calculations,  but there is a nonzero electric field
component $E_y$ (see above). Furthermore, we see that for opposite charge pairs 
the $sine$ and $cosine$ correlation terms are virtually identical, which, 
according to Refs.~\cite{BKL09,BKL10}, is hard to reconcile with a sizable 
elliptic flow in these collisions. However, the centrality distributions of 
opposite charge pions  exhibit contrary trends:
the  STAR measurement is positive and decreases but the HSD result
 is negative and increases toward central collisions where all components of 
angular correlations $\approx$0. It is noteworthy  that the UrQMD model shows 
quite close results. Indeed, $\langle \cos (\phi_\alpha-\phi_\beta) \rangle$ is just 
the sum of $cosine$ and $sine$ terms. So, summing the two opposite charge curves in 
Fig.~\ref{same-opp} we reproduce the UrQMD results presented in Fig.5
of Ref.~\cite{STAR-CME}. 

Recently  there were proposals to explain the observed CME effect without 
reference to the local parity violation. Different background
 mechanisms of the azimuthal correlation are considered: cluster decays, 
local transverse momentum conservation and local electric charge 
conservation~\cite{Wa09,Pr10,BKL09,BKL10}. These mechanisms may contribute 
to the effect under discussion, but their simplified  
estimates made in these papers are not able to describe the STAR measurements. 
Generally speaking, all these effects, such as the decay of resonances 
including heavy ones and the exact conservation of electric charge and 
energy-momentum,  are involved 
in our transport model but they do not help.  However, there is a 
conflicting point here.  The HSD model treats the system evolution for Au+Au(200 
GeV) in terms of hadrons and strings while the decisive phenomena  occur for times 
$t\lsim$0.3 fm/c. This is definitely in a non-equilibrium quark-gluon state.
On the other hand, as was shown in the multi-phase model~\cite{MZ11}, the charge
separation can be significantly reduced by the evolution of the quark-gluon plasma
produced in relativistic heavy-ion collisions and by the subsequent hadronization 
process. 

The conservation of the transverse momentum is a possible source of azimuthal 
correlations which was suggested to be a significant contribution to the measured
observable (Eq.~\ref{cos})~\cite{Pr10,BKL09,BKL10}). Using Eq.(\ref{cossin}) and 
the conservation of the transverse momenta, it was shown that 
roughly~\cite{Pr10,BKL09}) 
\begin{equation}
\label{cosv2}
\langle \cos (\phi_\alpha+\phi_\beta-2\Psi_{RP}) \rangle\simeq- \frac{v_2}{N}~,
\end{equation}
where $v_2$ is the elliptic flow coefficient measured for all produced particles and 
$N$ is the total number of all produced particles. It is definitely a qualitative 
result, but it demonstrates a close relation of the observed charge separation
effect with the elliptic flow. Equation (\ref{cosv2}) was specified more accurately 
in Ref.~\cite{BKL09}) and the correlator (\ref{cos}) was estimated under reasonable 
assumptions. It was concluded that transverse momentum conservation alone is not 
sufficient for explaining the STAR data. One should add that this issue has been 
considered in the  multi-phase model~\cite{MZ11} as well, with 
the conclusion that the charge particle separation
leads to a modification of the relation between the charge azimuthal correlation 
and the elliptic flow that is expected from transverse momentum conservation 
(Eq.~\ref{cosv2}). An essential point from this discussion  is that the charge
separation effect is roughly proportional to the elliptic flow; however, this 
quantity is underestimated in the HSD model resulting in a small value of the
correlator. We hope that future calculations  within the PHSD model~\cite{PHSD} 
might provide azimuthal correlations that will be closer to the measured data.
At present the   questions  noted above regarding the experimental observations
have no simple explanation.

\section{Summary and outlook}

We have extended the hadron string dynamics model to describe the formation of the
retarded electromagnetic field, its evolution during a nuclear collision and
the effect of this field on particle propagation. The case of the Au+Au collision at
$\sqrt{s_{NN}}$ for $b=$10 fm is considered in great detail. It is quite important
to understand the interplay of strong and electromagnetic interactions in this case
since it provides a point which is decisive  in the CME measurements at RHIC
as a function of centrality~\cite{Vol09,STAR-CME}.  It is shown that the most 
intensive magnetic field oriented perpendicularly to the reaction plane 
is formed during the
time when the Lorentz-contracted nuclei are passing through each other,
$t\lsim$0.2 fm/c. The maximal strength of the magnetic field here attains very
high values, $eB_y/m_\pi^2\sim$5.  This magnetic field strength is higher by
about 3-4 orders of magnitude than that at the surface of the 
magnetar~\cite{magnetar} which in turn is only slightly above the field strength 
in the star core~\cite{DT92}. Still larger magnetic fields up to $B\sim 10^{24}$ G 
might have appeared in the early universe~\cite{GR01}. It is impossible to make 
steady fields
stronger than $4.5\cdot 10^5$ gauss in the laboratory because the magnetic stresses
of such fields exceed the tensile strength of terrestrial materials.

This maximal strength of the magnetic field is created predominantly  by 
spectators. When target and projectile remainders   are separated, the 
spectator contribution goes
sharply down and for $t\sim$1 fm/c decreases by more than three orders
of magnitude. In subsequent times the participants come into the game, but their
contributions are small due to the mutual compensation of approximately equal number
of positive and negative charges as well as to the suppressive role of the
relativistic retardation effect.

The general pattern of the magnetic field is highly inhomogeneous. However, in
the ``almond'' transverse area (besides the boundary region) the field distributions
in $z$ or time look very similar, which allows us to use some simplifying assumptions
in phenomenological CME estimates~\cite{SIT09,TV10}.

The important accompanying quantity is the energy density $\varepsilon$ of the 
created particles. Its space-time distributions have been presented. 
It was shown that the
location of maxima in the field strength $eB_y$ and the energy density $\varepsilon$
nicely correlate with each other. Thus, it is a necessary condition for a realization
of the CME.

In the early time moments the created electric field is perpendicular to the magnetic
one and has a dominant $x$-component. In contrast to the $eB_y$ distribution,
the $eE_x$ distribution in the ($x-z)$ space plane has a minimum in the center of
the overlap region due to the ``hedgehog'' field structure of an isolated electric
charge.
The maximal strength of the electric field is by a factor about 5 lower than that
of the magnetic field. For $t\gsim$0.20 fm/c the electric field is small and can 
be even neglected.

The electromagnetic field is only moderately (within factor $\sim$5) changed 
with impact parameter and collision energy 
(for $\sqrt{s_{NN}}\approx$200-60 GeV) and strongly
suppressed for $t\gsim$0.2 fm/c. The scale  of the energy density change
is much larger but up to times of a few fm/c $\varepsilon >$1 GeV/fm$^3$ and
thereby  does not prevent quark-gluon plasma formation. Certainly, the issue 
of thermalization
remains open in this consideration. The very small influence of the 
electromagnetic field on observables is due to rather large masses of 
quasiparticles and the point that the system spends a very short time in a 
state with an extremely high electromagnetic  field. 

The comparison of global observables calculated in the HSD model with and without
an electromagnetic field reveals no difference apart from the transverse momentum
dependence of the elliptic flow where the model results slightly differ
in the range $p_{t}\sim$1 GeV/$c$.

Our analysis of the angular correlators - specific for the CME shows that
the calculated HSD background is very small and not able to describe the STAR
measurements~\cite{Vol09,STAR-CME}. The consideration of in-plane and out-of-plane 
projection components of this correlator does not allow 
us to clarify the picture and rises new questions related
to the experiment~\cite{BKL09,BKL10}. In this respect it is of great interest
to include quark-gluon degrees of freedom  directly in our approach. In particular,
the partonic generalization of the HSD model (PHSD)~\cite{PHSD} is highly suited for
this aim. Another way to approach the CME is to simulate an induced chromoelectric
field  which is assumed to be the source of the observed electric charge separation
of pion pairs relative to the reaction plane. Alternative mechanisms of
charge azimuthal asymmetry should be also carefully studied. Certainly, 
measurements of the CME at other bombarding energies as well as a search for new 
observables are very important.

\section*{Acknowledgments}
We are thankful to M. Gorenstein, D. Kharzeev, V. Koch, V. Skokov, and 
H. Warringa for useful discussion and remarks. This work has been supported 
by the LOEWE Center HIC for FAIR. V.T. is partially supported by a 
Heisenberg-Landau grant.

\end{document}